\documentclass[aps,notitlepage,twocolumn,amsmath,amssymb,amsfonts,showpacs,floats]{revtex4-1}

\usepackage{graphicx}
\usepackage{multirow}
\usepackage{color}
\usepackage{amsthm}
\usepackage{ulem}
\usepackage{enumerate}
\usepackage{rotating}
\usepackage[table]{xcolor}
\usepackage{hyperref} % should be at the end of all the other packages

\newcommand{\mc}[1]{\ensuremath{\mathcal{#1}}}
\newcommand{\mr}[1]{\ensuremath{\mathrm{#1}}}
\newcommand{\mbb}[1]{\ensuremath{\mathbb{#1}}}

\newcommand{\tr}{\ensuremath{\mathrm{Tr}}}

\newcommand{\ket}[1]{\ensuremath{| #1 \rangle}}
\newcommand{\prj}[1]{\ensuremath{| #1 \rangle \langle #1 |}}
\newcommand{\ovl}[2]{\ensuremath{\langle #1 | #2 \rangle}}
\newcommand{\epv}[2]{\ensuremath{\langle #1 | #2 | #1 \rangle}}
\newcommand{\matel}[3]{\ensuremath{\langle #1 | #2 | #3 \rangle}}

\newcommand{\ua}{\ensuremath{{\uparrow}}}
\newcommand{\da}{\ensuremath{{\downarrow}}}
\newcommand{\Ua}{\ensuremath{{\Uparrow}}}
\newcommand{\Da}{\ensuremath{{\Downarrow}}}

\begin{document}

\title{
Macroscopic Quantum Entanglement of a Kondo Cloud at Finite Temperature
}
\author{S.-S. B. Lee}%\email{ssblee@kaist.ac.kr}
\author{Jinhong Park}
\author{H.-S. Sim}\email[Corresponding author. ]{hssim@kaist.ac.kr}
\affiliation{Department of Physics, Korea Advanced Institute of Science and Technology, Daejeon 305-701, Korea}

\begin{abstract}
We propose a variational approach for computing the macroscopic entanglement in a many-body {\it mixed} state, based on entanglement witness operators, and compute the entanglement of formation (EoF), a mixed-state generalization of the entanglement entropy, in single- and two-channel Kondo systems at finite temperature. The thermal suppression of the EoF obeys power-law scaling at low temperature. The scaling exponent is halved from the single- to the two-channel system, which is attributed, using a bosonization method, to the non-Fermi liquid behavior of a Majorana fermion, a ``half'' of a complex fermion, emerging in the two-channel system. Moreover, the EoF characterizes the size and power-law tail of the Kondo screening cloud of the single-channel system.
\end{abstract}
%\date{\today}
\maketitle

Systems of many interacting particles often exhibit unusual macroscopic phenomena at zero temperature.
A useful concept of understanding their quantum nature is macroscopic entanglement~\cite{Amico08,Vedral08}, quantum correlation of many particles that cannot be imitated by classical correlations~\cite{Horodecki09}.
A popular measure for this purpose is entanglement entropy (EE). It captures entanglement between two macroscopic subsystems, and quantifies
new aspects of many-body ground states, including area law~\cite{Eisert10}, topological order~\cite{Kitaev06,Levin06}, and quantum criticality~\cite{Calabrese09,Vidal03}.

Generalizing this zero-temperature study is desirable, to explore how the macroscopic entanglement thermally decays or spatially extends.
This requires to study a {\it mixed} state, in which quantum and classical correlations coexist.
%For example, at finite temperature or by environments,
At finite temperature, a system is in a probabilistic mixture of energy eigenstates.
Its entanglement will reveal quantumness in quantum-to-classical crossover, collective excitations, decoherence, etc.
%It is worthwhile to study macroscopic entanglement at finite temperature, 
Moreover, EE measures entanglement only between two complementary subsystems in a pure state~\cite{Amico08}, providing  limited information about the spatial
extension of macroscopic entanglement. 
%More direct information is encoded in the entanglement between, e.g.,
To get more direct information, it is useful to consider, e.g., 
two distant {\it non-complementary} subsystems with changing the distance,
which 
%The two non-complementary subsystems 
are described by a mixed state, after
the remainder is traced out of a ground or thermal mixed state.

The computation of macroscopic entanglement in many-body mixed states, however, requires huge  costs.
For mixed states, EE unpredictably overestimates entanglement, since
it cannot distinguish between quantum and classical correlations.
Thus EE is
%cannot rule out classical correlation coexisting with quantum entanglement, thus it is 
generalized~\cite{Bennett96} into the entanglement of formation (EoF) $\mc{E}_\mr{F}$.
% via the convex-roof construction~\cite{Plenio07,Guhne09}; 
%for pure states, it equals the EoF. 
EoF quantifies the entanglement between two complementary subsystems A and B of a {\it mixed} state $\rho$ as
\begin{equation}
  \mc{E}_\mr{F} (\rho) = \inf_{\rho = \sum_i p_i \prj{\psi_i}} [ \sum_i p_i \, \mc{E}_\mr{E} (\ket{\psi_i}) ]. \label{ConvRoof}
\end{equation}
%According to Eq.~\eqref{ConvRoof}, 
It is obtained by exploring the possible decompositions of $\rho$ into normalized pure states $\ket{\psi_i}$ with weight $p_i$, 
%$\rho = \sum_i p_i \prj{\psi_i}$, 
and finding the optimal decomposition for which $\sum_i p_i \, \mc{E}_\mr{E} (\ket{\psi_i})$ is the lowest. Here, $\mc{E}_\mr{E} (\ket{\psi_i}) \equiv - \tr ( \rho_\mr{A} \log_2 \rho_\mr{A} )$ is the EE of $\ket{\psi_i}$ between A and B, 
% $\tr (\cdot)$ is the trace, 
and $\rho_\mr{A} = \tr_\mr{B} \prj{\psi_i}$ %; the state $\rho_\mr{A}$ of A 
is obtained from $\ket{\psi_i}$ by tracing out B.
For pure states $\ket{\psi}$, EoF reduces to EE, $\mc{E}_\mr{F}(\ket{\psi}) = \mc{E}_\mr{E}(\ket{\psi})$. The computational cost of exploring the decompositions is huge even for a small system of a three-qubit full-rank state, equivalent to that of minimizing a function of $63 \sim 959$ variables~\cite{Lee12,Ryu12,Roth_PRA}, and it exponentially increases with system size~\cite{Guhne09,Plenio}.
Most entanglement measures require such heavy costs~\cite{Plenio07,Guhne09}.
Negativity~\cite{Vidal02,Plenio05,Bayat10a,Bayat12,Calabrese12} is an exception,
however, cannot detect bound entanglement~\cite{Guhne09,Horodecki09}
that can appear in many-body systems \cite{Ferraro08,Santos12}. 
%
%since the number of variables scales as the cube of the rank of $\rho$.
Mutual information~\cite{Horodecki09} is not applicable to mixed-state entanglement, as it cannot distinguish between quantum and classical correlations.

On the other hand, in 
%the representative many-body problems of
Kondo effects~\cite{Hewson93}, 
the ground states have the entanglement between
the Kondo impurity spin and the surrounding conduction electrons, the latter forming 
%a macroscopic object called 
Kondo cloud~\cite{Affleck10,Mitchell11,Park13}.
Naturally, macroscopic entanglement would be a direct tool for characterizing the 
properties of the cloud, the essence of Kondo effects,
that cannot be captured by  few-particle correlations~\cite{Borda07,Holzner09}. %the correlations of a few particles.
For example, it will be meaningful to characterize the tail of the cloud by macroscopic entanglement, since the tail is expected to reflect the universality of low-energy Kondo physics in real space~\cite{Mitchell11}. Moreover, since Kondo effects have gapless excitations at any low temperature, it is important to study, by macroscopic entanglement, not only  their ground state but also their thermal suppression.
However,  the understanding of the macroscopic entanglement remains unsatisfactory due to the computation difficulty mentioned above, despite efforts~\cite{Sorensen07b,Eriksson11,Bayat10a,Bayat12}.

\begin{figure}
\centerline{\includegraphics[width=0.47\textwidth]{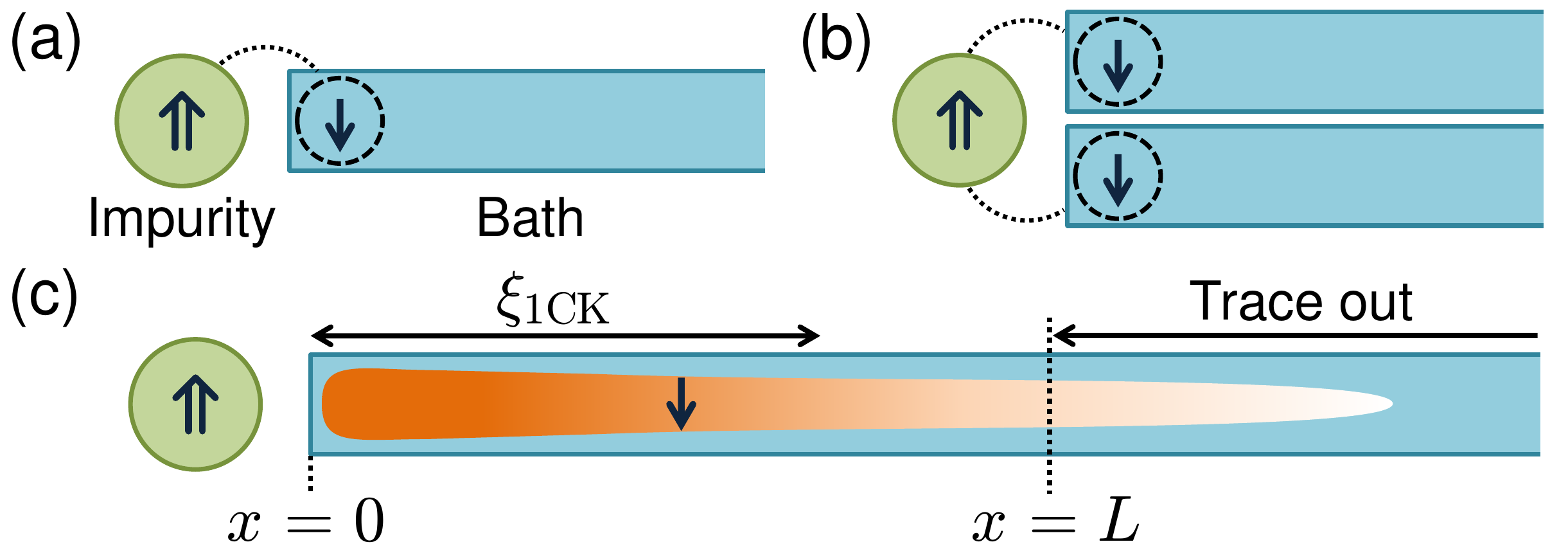}}
\caption{(Color Online)
%{Kondo systems and entanglement.}
(a) Single- (1CK) and (b) two-channel Kondo (2CK) systems.
In 1CK (2CK), a spin-1/2 impurity is antiferromagnetically coupled with the spin(s) of a single (two) channel(s) 
%(two channels) 
of a conduction electron bath at the impurity site~\cite{Hewson93,SM};
we consider a one-dimensional semi-infinite bath, without loss of generality.
(c) In 1CK, the Kondo cloud, a macroscopic electronic object of size $\xi_\mr{1CK}$, forms to screen the impurity spin at zero temperature.
%Here $\xi_\mr{1CK} = \hbar v_\mr{F} / k_\mr{B} T_\mr{K}$ and $v_\mr{F}$ is Fermi velocity.
The cloud spin ($\ua$, $\da$) entangles with the impurity spin ($\Ua, \Da$), forming the Kondo singlet of the Bell-state type $\ket{\Ua \da} - \ket{\Da \ua}$.
The entanglement of formation $\mc{E}_\mr{F}$ between the impurity at $x = 0$ 
and the bath electrons inside distance $L$ ($x \leq L$) quantifies how the macroscopic entanglement of the cloud spatially extends.
$\mc{E}_\mr{F}$ is reduced from the value of $L \to \infty$, if  the entanglement between electrons  outside $L$ (which are traced out) and the rest exists.
%We obtain the mixed quantum state $\rho$ by building thermal state at finite $T$ by numerical renormalization group (NRG) methods~\cite{Weichselbaum07,Bulla08} and then by tracing out the subsystem of $x > L$;
%we develop a way for the latter within NRG (see Eq.~\eqref{P_L}).
}
\label{fig_setup}
\end{figure}

In this Letter, we propose a variational approach for computing macroscopic entanglement in mixed states, based on entanglement witness operators (EWs)~\cite{Brandao05,Eisert07,Guhne09,Park10,Lee12,Ryu12}, and develop it for 
%the representative many-body effects of
single- (1CK) and two-channel Kondo (2CK) systems, using numerical renormalization group (NRG) methods~\cite{Weichselbaum07,Bulla08}.
We compute the EoF $\mc{E}_\mr{F}$ between the impurity and the electrons located within distance $L$ from the impurity at temperature $T$; see Fig.~\ref{fig_setup}. %The dependence of $\mc{E}_\mr{F}$ on $T$ unveils how the macroscopic entanglement thermally decays. In addition to the expected crossover around Kondo temperature $T_{1CK (2CK)}$ for the 1CK (2CK), it follows, at low temperature, the universal power laws of 
In addition to the expected crossover around the Kondo temperature $T_\mr{1CK (2CK)}$ of 1CK (2CK), 
the macroscopic entanglement measured by $\mc{E}_\mr{F}$ exhibits, at low temperature, the universal power-law thermal decay of $\mc{E}_\mr{F} \simeq 1 - a_{1} (T/T_\mr{1CK})^{2}$ for 1CK and $\mc{E}_\mr{F} \simeq 1 - a_{2} T/T_\mr{2CK}$ for 2CK; $a_{1, 2}$ are constants. 
The halving of the power-law exponent from 1CK to 2CK is attributed, using bosonization methods~\cite{Zarand00}, to a Majorana fermion
%, a ``half'' of a complex fermion, 
emerging in 2CK. Moreover, for 1CK, the dependence of $\mc{E}_\mr{F}$ on $L$ characterizes the spatial profile of the Kondo cloud. 
%; a similar idea was used in a recent proposal for experimentally detecting Kondo cloud~\cite{Park13}.  
The cloud size is $\xi_\mr{1CK} = \hbar v_\mr{F} / k_\mr{B} T_\mr{1CK}$ and robust against thermal effects at $T \lesssim T_\mr{1CK}$, while it decreases with increasing $T$ at $T \gtrsim  T_\mr{1CK}$; %roughly as $\hbar v_\mr{F} / (k_B T)$; 
$v_\mr{F}$ is Fermi velocity. At $T =0$, the cloud tail obeys another power law, $\mc{E}_\mr{F} \simeq 1- b_1 (\xi_\mr{1CK}/L)$; $b_1$ is a constant.
The different exponents of 1CK imply that the $T$ and $L$ dependences of $\mc{E}_\mr{F}$ have separate informations about entanglement in thermal states and the spatial extension of entanglement. %, respectively.

{\it Variational approach.---}
EWs are the physical operators %introduced to detect 
detecting whether a state is entangled~\cite{Guhne09,Horodecki09}.
They have been applied for quantifying entanglement in a few particles~\cite{Brandao05,Eisert07,Park10,Lee12,Ryu12}.
%It was suggested~\cite{Brandao05,Eisert07,Park10,Lee12,Ryu12} that they can be applied to the quantification;
Here we suggest to use EWs to efficiently compute {\it macroscopic} entanglement.

We introduce how to compute the EoF $\mc{E}_\mr{F} (\rho)$ of a target state $\rho$ by EW.
One finds the set $\mbb{M}_\rho$ of EWs $X$, whose expectation value provides a lower bound of $\mc{E}_\mr{F} (\rho)$ as  $\tr X \rho \le \mc{E}_\mr{F} (\rho)$. Here,
$\mbb{M}_\rho \equiv \{ X \, | \, \epv{\psi}{X} \leq \mc{E}_\mr{F} (\ket{\psi}), \,\, \forall \ket{\psi} \in \mc{H}_\rho \}$ and a Hilbert space $\mc{H}_\rho$ includes the range of $\rho$.
Among $X$'s, the optimal EW~\cite{Brandao05,Eisert07} of the largest expectation value provides $\mc{E}_\mr{F} (\rho)$,
\begin{equation}
  \mc{E}_\mr{F} (\rho) = \sup_{X \in \mbb{M}_\rho} \tr X \rho .
  \label{OW}
\end{equation}
It is equivalent to Eq.~\eqref{ConvRoof}, and the cost of exploring all operators in $\mbb{M}_\rho$ is huge.
Because of the difficulty in Eqs.~\eqref{ConvRoof} and ~\eqref{OW}, macroscopic entanglement in a thermal many-body state remains unexplored. % has not been quantified. 

%; but, it has  not been applied to quantify macroscopic entanglement

In our approach, instead of fully exploring $\mbb{M}_\rho$,
we construct an appropriate variational form of EW, which covers only a small subset of $\mbb{M}_\rho$ but includes or is close to the optimal EW.
Within the form, we find the operator $X^\mr{opt}_\rho$ whose expectation value $\tr X^\mr{opt}_\rho \rho$ is the largest.

A lower bound of $\mc{E}_\mr{F} (\rho)$ is obtained as $\tr X^\mr{opt}_\rho \rho$, because $X^\mr{opt}_\rho$ is an EW.
We obtain an upper bound by finding  a pure-state decomposition of $\rho$, based on the duality~\cite{Lee12,Ryu12}: 
%We obtain an upper bound, as below, by finding a pure-state decomposition of $\rho = \sum_i p'_i \prj{\psi'_i}$.
%The finding is based on the duality~\cite{Lee12,Ryu12} that
The optimal decomposition of $\rho$ in Eq.~\eqref{ConvRoof} is a mixture of the pure states in a set $\mbb{P}_{X} \equiv \{ \ket{\psi} \, | \, \epv{\psi}{X} = \mc{E}_\mr{F} (\ket{\psi}) \}$, if and only if  $\tr X \rho = \mc{E}_\mr{F} (\rho)$ in Eq.~\eqref{OW}.
We obtain $\mbb{P}_{X_\rho^\mr{opt}}$ 
and search a decomposition of $\rho = \sum_i p'_i \prj{\psi'_i}$,  each $\ket{\psi'_i}$ being sufficiently similar to an element of $\mbb{P}_{X^\mr{opt}_\rho}$~\cite{SM}.
Then, $\sum_i p'_i \, \mc{E}_\mr{F} (\ket{\psi'_i})$ is an upper bound.
The upper and lower bounds are close together (hence to the exact value) when $X^\mr{opt}_\rho$ is ``good''.
A good variational form can be constructed for a system at low temperature, considering  %the entanglement of 
its ground states and low-energy excitations, as shown below.

{\it EW in Kondo models.---}
We further develop this approach for Kondo systems; see Fig.~\ref{fig_setup}. Their Hamiltonian is $H = J \sum_\alpha \vec{S} \cdot \vec{s}_\alpha + \sum_{\alpha k \sigma} \epsilon_{k} c_{\alpha k \sigma}^\dagger c_{\alpha k \sigma}$. 
%, to compute the EoF $\mc{E}_\mr{F}$ between the Kondo impurity and the electrons within distance $L$ from the impurity at temperature $T$;
%The system Hamiltonian is written as
$J$ is the coupling strength between the impurity spin $\vec{S}$ and the electron spin $\vec{s}_\alpha = \sum_{k k' \sigma \sigma'} c^\dagger_{\alpha k \sigma} \vec{\sigma}_{\sigma \sigma'} c_{\alpha k' \sigma'}/2$ in channel $\alpha \in [1,M]$ at the impurity site ($x=0$), $M=1$ (2) for 1CK (2CK), $\vec{\sigma}$ is Pauli matrix, $c_{\alpha k \sigma}^\dagger$ creates an electron with spin $\sigma$, momentum $k$, and energy $\epsilon_k$ in channel $\alpha$~\cite{SM}.

To compute the EoF, we obtain the state $\rho$, by building thermal states by NRG~\cite{Weichselbaum07,Bulla08} and by tracing out the subsystem outside $L$. %of $x > L$;
We develop a way for the latter within NRG~\cite{SM}.
%in Eq.~\eqref{P_L} in Supplemental Material~\cite{SM}.
The resulting $\rho$ generally has rank $\sim 10^4$ too high to exactly obtain $\mathcal{E}_\mr{F}(\rho)$.

At $T = 0$ and $L \rightarrow \infty$, we exactly obtain optimal EWs based on
our derivation~\cite{SM} of the optimal EW $X_\mr{2qb}$ for a general two-qubit state $\rho_\mr{2qb}$, which provides the value of $\mc{E}_\mr{F} (\rho_\mr{2qb}) = \tr X_\mr{2qb} \rho_\mr{2qb}$ and satisfies $\epv{\psi}{X_\mr{2qb}} \leq \mc{E}_\mr{F} (\ket{\psi})$ for any pure state $\ket{\psi}$.
%our finding that, for any two-qubit state $\rho_\mr{2qb}$, there exists the optimal EW $X_\mr{2qb}$ which provides the value of $\mc{E}_\mr{F} (\rho_\mr{2qb}) = \tr X_\mr{2qb} \rho_\mr{2qb}$ and satisfies $\epv{\psi}{X_\mr{2qb}} \leq \mc{E}_\mr{F} (\ket{\psi})$ for any pure state $\ket{\psi}$; we derive $X_\mr{2qb}$ in Supplementary Material~\cite{SM}.
The 1CK ground state (so-called Kondo singlet), $\ket{G_\mr{1CK}} = \frac{1}{\sqrt{2}} ( \ket{\Ua} \ket{g_{-1/2}} - \ket{\Da} \ket{g_{1/2}})$, is a two-qubit Bell state of maximal entanglement ($\mc{E}_\mr{F} = 1$) between impurity spin states $\ket{\eta = \Ua, \Da}$ and bath states $\ket{g_{N_s}}$ of spin-$z$ quantum number $N_s$, satisfying $\ovl{g_{N_s}}{g_{N'_s}} = \delta_{N_s N'_s}$.
The optimal EW for $\mc{E}_\mr{F} (\ket{G_\mr{1CK}})$ has the form
\begin{equation}
    X_\mr{G1} = \frac{2}{\log 2} \prj{G_\mr{1CK}} - \left( \frac{2}{\log 2} - 1\right) I_\mr{G1},
  \label{X1}
\end{equation}
where $I_\mr{G1} = \sum_{\eta = \Ua,\Da, \, N_s = \pm 1/2} \prj{\eta} \otimes \prj{g_{N_s}}$ is the identity operator of the two-qubit Hilbert subspace for $\ket{G_\mr{1CK}}$.
Notice $\epv{G_\mr{1CK}}{X_\mr{G1}} = \mc{E}_\mr{F} (\ket{G_\mr{1CK}}) = 1$.

The two-fold degenerate ground states of 2CK are also two-qubit Bell states ($\mc{E}_\mr{F} = 1$),
$\ket{G_\mr{2CK}^+} = \frac{1}{\sqrt{2}} (\ket{\Ua} \ket{g_{0}^+} + \ket{\Da} \ket{g_{1}^+})$ and
$\ket{G_\mr{2CK}^-} = \frac{1}{\sqrt{2}} (\ket{\Ua} \ket{g_{-1}^-} + \ket{\Da} \ket{g_{0}^-})$,
where $\ket{G_\mr{2CK}^\pm}$ has the total spin-$z$ quantum number of $\pm 1/2$ and $\ket{g^\pm_{N_s}}$ is the bath state associated with $\ket{G_\mr{2CK}^\pm}$, satisfying $\ovl{g_{N_s}^p}{g_{N'_s}^{p'}} = \delta_{p p'} \delta_{N_s N'_s}$.
Thus the 2CK state at $T = 0$ and $L \rightarrow \infty$ is $\rho_\mr{G2} = (\prj{G_\mr{2CK}^+} + \prj{G_\mr{2CK}^-})/2$.
The optimal EW $X_\mr{G2} = X_\mr{G2+} + X_\mr{G2-}$ provides the value of $\mc{E}_\mr{F} (\rho_\mr{G2}) = \tr X_\mr{G2} \rho_\mr{G2} = 1$,
where $X_\mr{G2\pm} = [2 \prj{G_\mr{2CK}^\pm} - ( 2 - \log 2 ) I_\mr{G2\pm} ] /\log 2$
%each quantify $\mc{E}_\mr{F} (\ket{G_\mr{2CK}^\pm}) = 1$
and $I_\mr{G2+}$ ($I_\mr{G2-}$) is the identity of two-qubit subspace $\mc{H}_\mr{G2+} = \{ \ket{\Ua}, \ket{\Da} \} \otimes \{ \ket{g_0^+}, \ket{g_1^+} \}$ ($\mc{H}_\mr{G2-} = \{ \ket{\Ua}, \ket{\Da} \} \otimes \{ \ket{g_{-1}^-}, \ket{g_0^-} \}$).
%Notice $\epv{G_\mr{2CK}^+}{X_\mr{G2+}} = \epv{G_\mr{2CK}^-}{X_\mr{G2-}} = \mc{E}_\mr{F} (\ket{G_\mr{2CK}^+}) = \mc{E}_\mr{F} (\ket{G_\mr{2CK}^-}) = 1$.

%To find an efficient variational form of $X$, we observe the following:
%Without loss of generality, here we define $\ket{g_{1/2}}$ to make the relative sign between $\ket{\Ua} \ket{g_{-1/2}}$ and $\ket{\Da} \ket{g_{1/2}}$ be $+$, rather than a common choice $-$ for singlets.
%We find such $X_\mr{2qb}$ by first constructing the EW for the concurrence~\cite{Wootters98} of $\rho_\mr{2qb}$~\cite{Park10} and then using the relation between the EoF and the concurrence; see Supplementary Material~\cite{SM} for detail.

For the state $\rho$ at general $T$ and $L$, we construct an EW $X$ variationally, generalizing $X_\mr{G1}$ and $X_\mr{G2}$, as follows.
(i) Decompose the whole Hilbert space into two-qubit subspaces $\mc{H}_i = \{ \ket{\Ua}, \ket{\Da} \} \otimes \{\ket{\phi_{i \Ua}}, \ket{\phi_{i \Da}} \}$, where $\{ \ket{\phi_{i\eta}} \}$ is an orthonormal basis of bath states.
We parametrize $\ket{\phi_{i\eta}}$'s for the optimization discussed below.
(ii) For each subspace $\mc{H}_i$, we obtain the optimal EW $X_i$ which provides $\mc{E}_\mr{F}(\rho_i) = \tr X_i \rho_i$. Here $\rho_i = I_i \rho I_i$ is the projection of $\rho$ onto $\mc{H}_i$ and $I_i$ is the  identity of $\mc{H}_i$.
This construction of $X_i$ depends on the choice of  $\{ \ket{\phi_{i\eta}} \}$. 
(iii) The sum $X = \sum_i X_i$ of the two-qubit EWs is our variational form. We optimize the choice of $\{ \ket{\phi_{i\eta}} \}$ (hence $X$), to make the lower and upper bounds of $\mc{E}_\mr{F} (\rho)$ closer~\cite{SM}.

For example, at $T \ll T_\mr{1CK,2CK}$ and $L \to \infty$, $\rho$ (for any of 1CK and 2CK) is a mixture of energy eigenstates $\ket{E_i}$ with energy $E_i \ll k_\mr{B} T_\mr{1CK,2CK}$.
%The eigenstates 
$\ket{E_i}  = b_{i \Ua} \ket{\Ua} \ket{e_{i\Ua}} + b_{i \Da} \ket{\Da} \ket{e_{i\Da}}$ has an analogous form ($\mc{E}_\mr{F} \lesssim 1$) to the Bell state, where $b_{i\eta} \simeq 1 / \sqrt{2}$, $\ovl{e_{i\eta}}{e_{i\eta'}} = \delta_{\eta\eta'}$, and $\ovl{e_{i\eta}}{e_{i' \neq i, \eta'}} \simeq 0$.
Hence we choose $\{ \ket{\phi_{i\eta}} \}$ by orthonormalizing $\{ \ket{e_{i\eta}} \}$ and construct $X_i$'s similarly to Eq.~\eqref{X1}.

%\red{(iii) The sum $X = \sum_i X_i$ of the two-qubit EWs is our variational form, providing a lower bound of $\mc{E}_\mr{F} (\rho)$ as $\tr (\sum_i X_i \rho)$.}
A lower bound of $\mc{E}_\mr{F} (\rho)$ is obtained as $\tr (\sum_i X_i \rho)$.
A upper bound is $\sum_j p_j' \mc{E}_\mr{F} (\ket{\psi_j'})$, by finding $\ket{\psi_j'}$ which is similar to a state in $\mbb{P}_{X}$ and satisfies $\sum_j p_j' \prj{\psi'_j} = \rho$; to avoid the huge cost of obtaining $\mbb{P}_{X}$, we use a subset $\bigcup_i \mbb{P}_{X_i} \subset \mbb{P}_{X}$.
To get better bounds, we optimize the choice of $\{ \ket{\phi_{i \eta}} \}$ and $\{ p'_j, \ket{\psi_j'} \}$, based on the structure of $\rho$~\cite{SM}. 
%To reduce the computation cost of finding $\ket{\psi_j'}$, we repeat the whole steps of finding $X_i$'s for each NRG block of $\rho$.
%We apply this strategy to general $T$ and $L$, and also to the analytic study of $\mc{E}_\mr{F} (\rho)$, as shown below.
%Note that 
$X$ cannot detect off-diagonal blocks $I_i \rho I_{i' \ne i}$ which are however made small at $T \ll T_\mr{1CK}$ and $L \gg \xi_\mr{1CK}$ by appropriately choosing $\{ \ket{\phi_{i \eta}} \}$.
We emphasize that the decomposition into the two-qubit subspaces allows us to avoid the impractical cost of computing EoF by Eq.~\eqref{ConvRoof}.

\begin{figure}
\centerline{\includegraphics[width=0.47\textwidth]{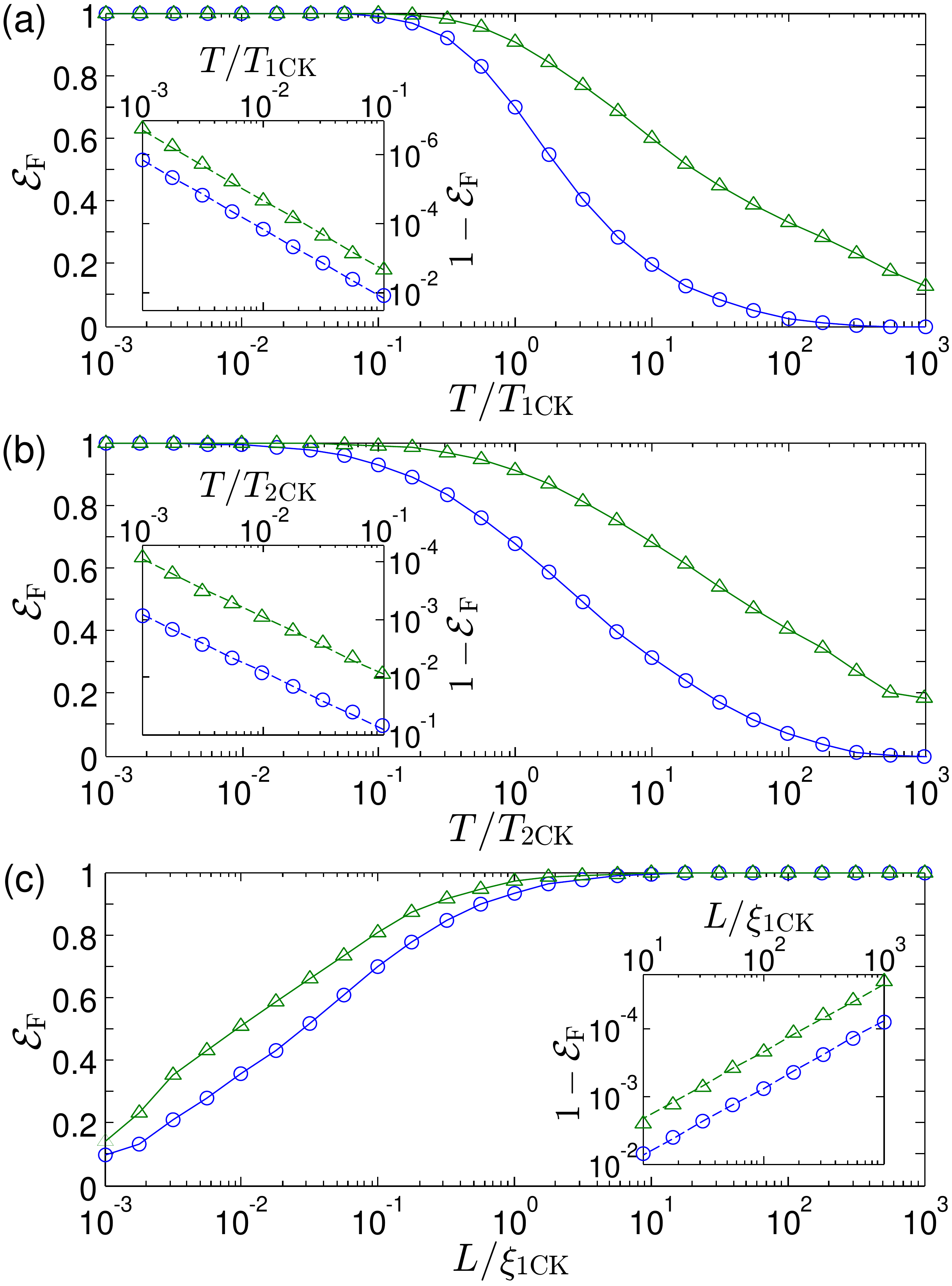}}
\caption{(Color Online)
Entanglement of formation $\mc{E}_\mr{F}$ between the Kondo impurity and the bath electrons inside $L$ at temperature $T$ for single- (1CK) and two-channel Kondo (2CK) systems.
(a,b) $\mc{E}_\mr{F}$ versus $T$ at $L \rightarrow \infty$ for (a) 1CK and (b) 2CK.
(c) $\mc{E}_\mr{F}$ versus $L$ at $T = 0$ for 1CK.
Blue circles (green triangles) denote the lower (upper) bounds of $\mc{E}_\mr{F}$.
These bounds are close to each other, especially at $T < T_\mr{1CK,2CK}$ and $L > \xi_\mr{1CK}$, enough to predict scaling behavior.
Insets: The universal scaling behavior of $\mc{E}_\mr{F}$ versus $T/T_\mr{1CK,2CK} \ll 1$ or $L/\xi_\mr{1CK} \gg 1$, well fitted by power laws (dashed lines); see Eqs.~\eqref{K_T} and \eqref{1CK_L}.
The NRG parameters used for this plot and the expression of $T_\mr{1CK,2CK}$ are given in Ref.~\cite{SM}. %
% We choose $J = 0.3$, $\Lambda = 4$ and $z = 0, 0.5$. The Kondo screening length is $\xi_\mr{K} = \hbar v_\mr{F} / k_\mr{B} T_\mr{K}$. $k_\mr{B} T_\mr{K} = D \sqrt{\nu J} e^{-1/\nu J}$. Here $T_\mr{K}$ for 2CK is numerically determined by the convergence of the lowest-energy eigenvalues.
}
\label{fig_res}
\end{figure}

{\it Result.---}
We discuss the result of the temperature dependence of $\mc{E}_\mr{F}$ at $L \to \infty$ in Fig.~\ref{fig_res}.
In both 1CK and 2CK, $\mc{E}_\mr{F}$ shows maximal entanglement at $T=0$, slowly decays with $T \lesssim T_\mr{1CK,2CK}$, and rapidly vanishes  at $T \gtrsim T_\mr{1CK,2CK}$, exhibiting the crossover around $T_\mr{1CK,2CK}$.
At $T \ll T_\mr{1CK,2CK}$, the upper and lower bounds show the same universal power-law decay in each system,
\begin{equation}
  \begin{alignedat}{2}
    \mc{E}_\mr{F} & \simeq 1 - a_1 (T/T_\mr{1CK})^2 & \qquad & \text{(1CK)}, \\
    \mc{E}_\mr{F} & \simeq 1 - a_2 (T/T_\mr{2CK}) & \qquad & \text{(2CK)}.
  \end{alignedat}
  \label{K_T}
\end{equation}
%It implies that 2CK has more fragile macroscopic entanglement at scaled low temperature.  

In Eq.~\eqref{K_T}, the scaling exponent 
is halved from 1CK to 2CK,
reflecting different low-energy excitations.
At $T \ll T_\mr{1CK,2CK}$, the thermal state $\rho = \sum_i w_i \prj{E_i}$ is governed by $\ket{E_i}$'s with $E_i \simeq k_\mr{B} T$, because of the competition between Boltzmann weight $w_i$ and degeneracy.
There are two sources suppressing $\mc{E}_\mr{F} (\rho)$:
(i) Each $\ket{E_i}$ is less entangled; %pure state
$\mc{E}_\mr{F} (\ket{E_i}) \simeq 1 - 2 |S_{z, ii}|^2 /\log 2$ for $E_i \ll k_\mr{B} T_\mr{1CK,2CK}$, where $S_z$ is the impurity spin-$z$ operator and $S_{z, i i} = \langle E_i |S_z | E_i \rangle = (|b_{i \Ua}|^2 - |b_{i \Da}|^2)/2$.
(ii) EoF satisfies the convexity, $\mc{E}_\mr{F} ( \sum_i w_i \prj{E_i} ) \le \sum_i w_i \mc{E}_\mr{F} (\ket{E_i})$.
Using bosonization~\cite{Zarand00}, we find that these two sources give the same exponent~\cite{SM}.
Here we explain the former factor.
A pseudofermion operator $c^{(\dagger)}$ describes the impurity spin as $S_z = c^\dagger c - 1/2$.
In 1CK, both $c^{\dagger}$ and $c$ couple to the bath~\cite{Zarand00}.
Since the coupling is energy dependent, each of $c^{\dagger}$ and $c$ gives a scaling factor $\sim \sqrt{T/T_{\mr{1CK}}}$,
leading to $1 - \mc{E}_\mr{F} (\ket{E_i}) \sim | S_{z,ii} |^2 \sim (T / T_\mr{1CK})^2$.
In 2CK, $S_z$ is rewritten as $S_z = i \gamma_+ \gamma_-$. Here,
only a Majorana fermion $\gamma_-$, a ``half'' of $c^{(\dagger)}$, couples to the bath, providing the factor $\sim \sqrt{T/T_{\mr{2CK}}}$;
the other Majorana $\gamma_+$ is decoupled, not giving $T$ independence.
This causes the exponent halving in 2CK. It is non-Fermi liquid behavior.
%Note a similar reduction of power-law exponents was reported in transport experiments~\cite{Mebrahtu13}.

The dependence of $\mc{E}_\mr{F}$ on $L$ is obtained for 1CK in Fig.~\ref{fig_res}(c). 
$\mc{E}_\mr{F}(L\to \infty) - \mc{E}_\mr{F}(L)$ indicates
entanglement between $x>L$ and the rest. %, which is lost in $\mc{E}_\mr{F}(L)$.
%; a proposal for Kondo cloud detection has similar motivation~\cite{Park13}.
At $T=0$, $\mc{E}_\mr{F}=1$ at $L \to \infty$ and decreases only slightly at $L > \xi_\mr{1CK}$, implying that Kondo cloud lies mostly (more than 90 \%) inside $\xi_\mr{1CK}$. The cloud has a long tail of the power law at $L \gg \xi_\mr{1CK}$,
\begin{equation}
  \mc{E}_\mr{F} \simeq 1 - b_1 ( \xi_\mr{1CK} / L ) \qquad \text{(1CK)}, \label{1CK_L}
\end{equation}
which is reproduced~\cite{SM} with Yosida's ground state~\cite{Yosida66}.
%We obtain the same power law~\cite{SM} from Yosida's ground state~\cite{Yosida66}.
%; see Eq.~\eqref{X_Y} in Supplemental Material~\cite{SM}.

At finite $T$, the $L$ dependence of $\mc{E}_\mr{F}$ characterizes the thermal reduction of Kondo cloud. We find that
the cloud size, within which the majority of the cloud lies, is $\xi_\mr{1CK}$ (almost insensitive to $T$) at $T \lesssim T_\mr{1CK}$, and decreases with $T$ at $T \gtrsim  T_\mr{1CK}$~\cite{SM}.
%roughly as $\hbar v_\mr{F} / (k_B T)$; 
%(Fig.~\ref{fig_cloud} in Supplemental Material~\cite{SM}).
Moreover, the two 1CK power laws in Eqs.~\eqref{K_T} and \eqref{1CK_L} have different exponent, not connected by $L \leftrightarrow \hbar v_\mr{F} / k_\mr{B} T$ from the uncertainty principle, and they are additive at $L \gg \xi_\mr{1CK}$ and $T \ll T_\mr{1CK}$ as $1 - \mc{E}_\mr{F} \simeq a_{1} (T/T_\mr{1CK})^2 + b_{1} (\xi_\mr{1CK}/L)$~\cite{SM}.
%(Fig.~\ref{fig_cloud2} in Supplemental Material~\cite{SM}).
These unusual results imply that entanglement suppression by thermal effects has different mechanism from that by the partial trace over $x > L$. 
%the subsystem outside distance $L$ from the impurity.
The former reflects thermal entanglement suppression, while the latter measures the spatial extension of entanglement.
Note that a mixed state obtained from a ground state by tracing out its subsystem is different from a thermal state, when the subsystem does not behave as a legitimate heat bath.

Finally, in contrast to EoF, correlations between the impurity spin and a conduction electron spin at $L$ do not detect macroscopic entanglement, because of the entanglement monogamy~\cite{Horodecki09} that tracing out all bath electrons except the one at $L$ leaves only negligible entanglement. They measure the cloud tail differently from $\mc{E}_\mr{F}$;
the spin-spin correlation~\cite{Borda07} decays as $1/L^2$ at $L \gg \xi_\mr{1CK}$, and the concurrence does not detect the cloud~\cite{Oh}.
%does not detect Kondo cloud~\cite{Oh}.
%Impurity entanglement entropy~\cite{Sorensen07b,Eriksson11} decays as $\xi_\mr{1CK}/L$ at $L \gg \xi_\mr{1CK}$ as in Eq.~\eqref{1CK_L}, but as $T/T_\mr{1CK}$ at $T \ll T_\mr{1CK}$, contrary to Eq.~\eqref{K_T}.
Impurity entanglement entropy~\cite{Sorensen07b,Eriksson11} detects macroscopic correlations, but it is not an entanglement measure; it decays as $\xi_\mr{1CK}/L$ at $L \gg \xi_\mr{1CK}$ as in Eq.~\eqref{1CK_L}, but as $T/T_\mr{1CK}$ at $T \ll T_\mr{1CK}$, contrary to Eq.~\eqref{K_T}.
Note that the cloud size at zero temperature was discussed in spin-chain Kondo models, using negativity~\cite{Bayat10a,Bayat12}. 

{\it Perspective.---}
We have proposed a viable approach for computing macroscopic entanglement in thermal mixed states.
%The macroscopic entanglement measured by EoF unveils the thermal decay and spatial profile of Kondo cloud, the essence of Kondo effects, and characterizes the non-Fermi liquid behavior of 2CK by Majorana fermions.
Our study implies that EoF is a good tool for quantifying macroscopic quantumness in many-body mixed states; its original operational meaning~\cite{Horodecki09} is a nonregularized entanglement cost in quantum information.

Our results indicate that
the macroscopic entanglement characterizes the new aspects of many-body systems at finite temperature, inaccessible by conventional means and by EE.
For example, it can identify the spatial extension of quantum correlations, the competition between the coexisting quantum and classical correlations induced by thermal effects or environments, and the fate of the zero-temperature correlations (e.g., topological order and quantum criticality) at finite temperature.

Our approach is optimized for computing entanglement between a few impurities and a macroscopic subsystem, and
directly applicable to quantum impurity problems.
It is in principle applicable to any convex-roof measures~\cite{Plenio07,Guhne09}, including multipartite entanglement~\cite{Lee12,Ryu12}, and useful for experimental entanglement detection~\cite{Park10,Guhne09}.
It is desirable to extend our approach to study entanglement between macroscopic subsystems.

Experimental evidence of Kondo cloud remains elusive~\cite{Affleck10,Park13}. It may be because the cloud is a macroscopic  object entangled with an impurity, showing rapid quantum fluctuations with zero average spin.
It will be valuable to find experimentally accessible EWs,
to confirm the entanglement, hence, the cloud.

%On the other hand,
%EWs are useful for experimental entanglement detection~\cite{Park10,Guhne09} and thus may provide a way to detect the Kondo cloud in experiment.
%The macroscopicity of Kondo cloud has been an obstacle in finding experimental evidence of Kondo cloud \cite{Affleck10,Park13}.
%%Kondo cloud is macroscopic in size and particle numbers;
%If one can construct EWs by combining experimentally accessible quantities, the entanglement would be measurable and directly indicate the Kondo cloud.

%that remains unexplored due to computation difficulty.

%Our approach may open a way of studying macroscopic entanglement in many-body mixed states at finite temperature, 

%Our approach may open a way of studying entanglement in macroscopic mixed states, especially in quantum impurity systems, that remains unexplored due to computation difficulty.

%We have proposed a viable approach for computing macroscopic entanglement in a mixed state at finite temperature.

%To our knowledge, it is the first time that macroscopic entanglement has been computed and analyzed for a thermal many-body  condensed-matter system. 

 %This suggests that macroscopic entanglement is a direct tool for characterizing the competition between the coexisting quantum and classical correlations induced by thermal effects or environments,  and  correlations among multiple subsystems in a ground state. 

We thank Ehud Altman, Henrik Johannesson, and Jan von Delft for valuable discussions,
Yong Hyun Kim for allowing us to use cluster computers in his group,
and the support by Korea NRF (Grant No. 2013R1A2A2A01007327).
%Grant No. 2011-0022955

%\section*{Methods}

%\section*{Acknowledgements}
%We thank Ehud Altman, Henrik Johannesson, and Jan von Delft for valuable discussions,
%Yong Hyun Kim for allowing us to use cluster computers in his group,
%and the support by Korea NRF (Grant No. 2011-0022955 and Grant No. 2013R1A2A2A01007327).
%
%\section*{Author contributions}
%S.-S.B.L. performed the numerical computation of EoF.
%S.-S.B.L. and J.P. performed the analytical calculations based on the bosonization.
%S.-S.B.L. and H.-S.S. developed the approach for computing entanglement, and wrote the manuscript.
%H.-S.S. supervised the project.
%
%\section*{Competing financial interests}
%The authors declare no competing financial interests.

\clearpage

\begin{widetext}

\begin{center}
{\large \bf Supplementary Material for

``Macroscopic quantum entanglement of Kondo cloud at finite temperature''}
\end{center}

% 수식과 그림 숫자를 리셋. 따로 .tex 파일을 만들 경우엔 필요없음
\setcounter{equation}{0}
\setcounter{figure}{0}

% 번호를 S1, S2, ... 로 붙게 함
\renewcommand{\theequation}{S\arabic{equation}}
\renewcommand{\thefigure}{S\arabic{figure}}

Here we provide the details of our approaches and some supplementary results.
In Sec.~\ref{KH}, we briefly introduce the Kondo models.
In Sec.~\ref{NRG}, we describe how to numerically construct the thermal mixed states of the Kondo models by NRG, and give the NRG parameters.
In Sec.~\ref{TO}, we describe in details the way of tracing out the subsystem in $x > L$, within the NRG formalism.
In Sec.~\ref{XI}, we derive the ``two-qubit'' EW $X_\mr{2qb}$.
In Sec.~\ref{WO}, we prove that the sum $X = \sum_i X_i$ is a valid EW, and discuss how to choose the bath basis $\{ \ket{\phi_{i\eta}} \}_{i\eta}$.
In Sec.~\ref{UB}, we give the way to obtain the upper bound of $\mc{E}_\mr{F} (\rho)$.
In Sec.~\ref{TB}, we analyze the scaling behavior of the thermal suppression of $\mc{E}_\mr{F}$ in Eq.~(4) in the main text, using the finite-size bosonization method.
In Sec.~\ref{L1CK}, we reproduce the long-tail scaling of $\mc{E}_\mr{F}(L)$ in Eq.~(5) in the main text, using the Yosida's variational ground state.
In Sec.~\ref{TL}, we give the computation result of EoF for 1CK when both $T$ and $L$ are finite, to discuss the size of the Kondo cloud at finite $T$.
We also address that, at $T \ll T_\mr{1CK}$ and $L \gg \xi_\mr{1CK}$, two power-law decays are additive.

\section{Kondo Hamiltonian}
\label{KH}

In 1CK (2CK), a spin-1/2 impurity is antiferromagnetically coupled with the spin(s) of a single channel (two channels) of the conduction electron bath at the impurity site~\cite{Hewson93}.
Without loss of generality, we consider a semi-infinite one-dimensional bath ranging from $x = 0$ (the impurity site) to $x \to \infty$.
Its Hamiltonian is
\begin{equation}
  H = J \sum_\alpha \vec{S} \cdot \vec{s}_\alpha + \sum_{\alpha k \sigma} \epsilon_{k} c_{\alpha k \sigma}^\dagger c_{\alpha k \sigma}, \label{KondoHamiltonian}
  \end{equation}
where $J$ is the coupling strength, $\vec{S}$ is the impurity spin operator, $\vec{s}_\alpha = \sum_{k k' \sigma \sigma'} c^\dagger_{\alpha k \sigma} \vec{\sigma}_{\sigma \sigma'} c_{\alpha k' \sigma'}/2$ is the electron spin operator in channel $\alpha \in [1,M]$ at the impurity site ($x=0$), $M=1$ (2) for 1CK (2CK), $\vec{\sigma}$ is Pauli matrix, $c_{\alpha k \sigma}^\dagger$ creates an electron with spin $\sigma$, momentum $k$, and energy $\epsilon_k = \hbar v_\mr{F} (k-k_\mr{F}) \in (-D,D)$ in $\alpha$, constant density of states $\nu=1/2D$, $k_\mr{F}$ is Fermi momentum, and $D$ is the bandwidth. In this work, we consider the following case: The two channels of 2CK have the same coupling strength, and there is no external magnetic field.
We use 1CK Kondo temperature $k_\mr{B} T_\mr{1CK} = D \sqrt{\nu J} e^{-1/\nu J}$, and 
%and Kondo cloud length as $\xi_\mr{1CK} = \hbar v_\mr{F} / k_\mr{B} T_\mr{1CK}$. 
determine $T_\mr{2CK}$ from energy-eigenvalue convergence in NRG.
% of the lowest-energy eigenvalues to the fixed point. 

%\section*{S2. Density operator by NRG}
\section{Density matrix by NRG}
\label{NRG}

In NRG~\cite{Bulla08},
each channel is logarithmically discretized and mapped onto a tight-binding chain (so-called Wilson chain) of length $N$, whose Hamiltonian is 
$ H_N = J \vec{S} \cdot \sum_{\alpha\sigma \sigma' } f_{\alpha 0 \sigma}^\dagger \frac{\vec{\sigma}_{\sigma \sigma'}}{2} f_{\alpha 0 \sigma'} +
     \sum_{n = 0}^N \sum_{\alpha \sigma} \left( t_n f_{\alpha n \sigma}^\dagger f_{\alpha n+1 \sigma} + \text{H.c.} \right)$,
%\begin{equation}
%    H_N = J \vec{S} \cdot \sum_{\alpha\sigma \sigma' } f_{\alpha 0 \sigma}^\dagger \frac{\vec{\sigma}_{\sigma \sigma'}}{2} f_{\alpha 0 \sigma'} +  \sum_{n = 0}^N \sum_{\alpha \sigma} \left( t_n f_{\alpha n \sigma}^\dagger f_{\alpha n+1 \sigma} + \text{H.c.} \right), \nonumber \end{equation}
where $f_{\alpha n \sigma}^\dagger$ creates an electron in the single-particle state $\ket{\alpha n \sigma}$ of spin $\sigma$ in channel $\alpha$ at site $n$, $t_n \sim D \Lambda^{-n/2}$, %is the hopping energy, 
and $\Lambda$ is the discretization parameter. $\ket{\alpha n \sigma}$ has energy $\sim D \Lambda^{-n/2}$ and extends over length $\sim k_\mr{F}^{-1} \Lambda^{n/2}$. $H_N$ is iteratively diagonalized, based on the energy-scale hierarchy. At each ($n$-th) iteration step, only the lowest-lying energy eigenstates $\{ \ket{E_{ni}^\mr{K}} \}_i$ of the Hamiltonian of the step are kept to construct the next-step Hamiltonian, while the rest $\{ \ket{E_{ni}^\mr{D}} \}_i$ is discarded. The discarded states are the energy eigenstates of $H_N$,  %\cite{Anders05}
$ H_N \ket{E_{ni}^\mr{D}} \otimes \bigotimes_{\alpha; n' > n} \ket{s_{\alpha n'}} \approx E_{ni}^\mr{D} \ket{E_{ni}^\mr{D}} \otimes \bigotimes_{\alpha; n' > n} \ket{s_{\alpha n'}}$,
%\begin{equation}
%  H_N \ket{E_{ni}^\mr{D}} \otimes \bigotimes_{\alpha; n' > n} \ket{s_{\alpha n'}} \approx E_{ni}^\mr{D} \ket{E_{ni}^\mr{D}} \otimes \bigotimes_{\alpha; n' > n} \ket{s_{\alpha n'}}, \nonumber \end{equation}
where $\ket{s_{\alpha n} = 0, \ua, \da, \ua\da}$ denote the occupation basis states of site $n$ and channel $\alpha$. %; this is the approximation of NRG. 

The equilibrium state at temperature $T$ is constructed~\cite{Weichselbaum07} as 
\begin{equation}
%  \begin{gathered}
    \rho (T) = \sum_n \rho_n \otimes I_{>n}, \quad
  \rho_n = \sum_i \frac{e^{-E^\mr{D}_{ni} / k_\mr{B} T}}{Z} \prj{E^\mr{D}_{ni}} , \quad
  I_{>n} = \bigotimes_{\alpha; n' > n} \sum_s \prj{s_{\alpha n'}},
%\end{gathered}
  \label{rhoT}
\end{equation}
where $k_\mr{B}$ is Boltzmann constant and $Z$ is the partition function. % and $\tr \rho = 1$.
Each block $\rho_n$ covers energy $\sim D \Lambda^{-n/2}$ and length $\sim k_\mr{F}^{-1} \Lambda^{n/2}$.
$\tr (\rho_n \otimes I_{> n})$ 
is maximal near $n = n_T \equiv -2 \log_\Lambda (k_\mr{B} T / D)$ due to the competition between Boltzmann factor $e^{ - D \Lambda^{-n/2} / k_\mr{B} T}$ and degeneracy $\tr \, I_{>n} = 4^{N-n}$.
In this work, we choose $\Lambda = 4$, $J/D = 0.3$, and the number of kept states $\lesssim 300$ at each iteration, and use the $z$-averaging~\cite{Bulla08} with $z=0$ and 0.5.
%$N$ large enough to achieve small enough energy resolution and to cover large enough system size.

%As each block $\rho_n$ involves restricted number of chain sites ($n+1$ sites and the impurity) and is decoupled from the rest,
%
%If all the blocks involve the same $N + 1$ sites, the Hilbert space for these blocks would increase exponentially with $N$. However, the fact that different number of sites are involved in $\rho_n$ indicates the hierarchical structure of the Hilbert space; thus the Hilbert space and $\rho (T)$ should be treated carefully.

\begin{figure}[bt]
\begin{center}
  \includegraphics[width=0.7\textwidth]{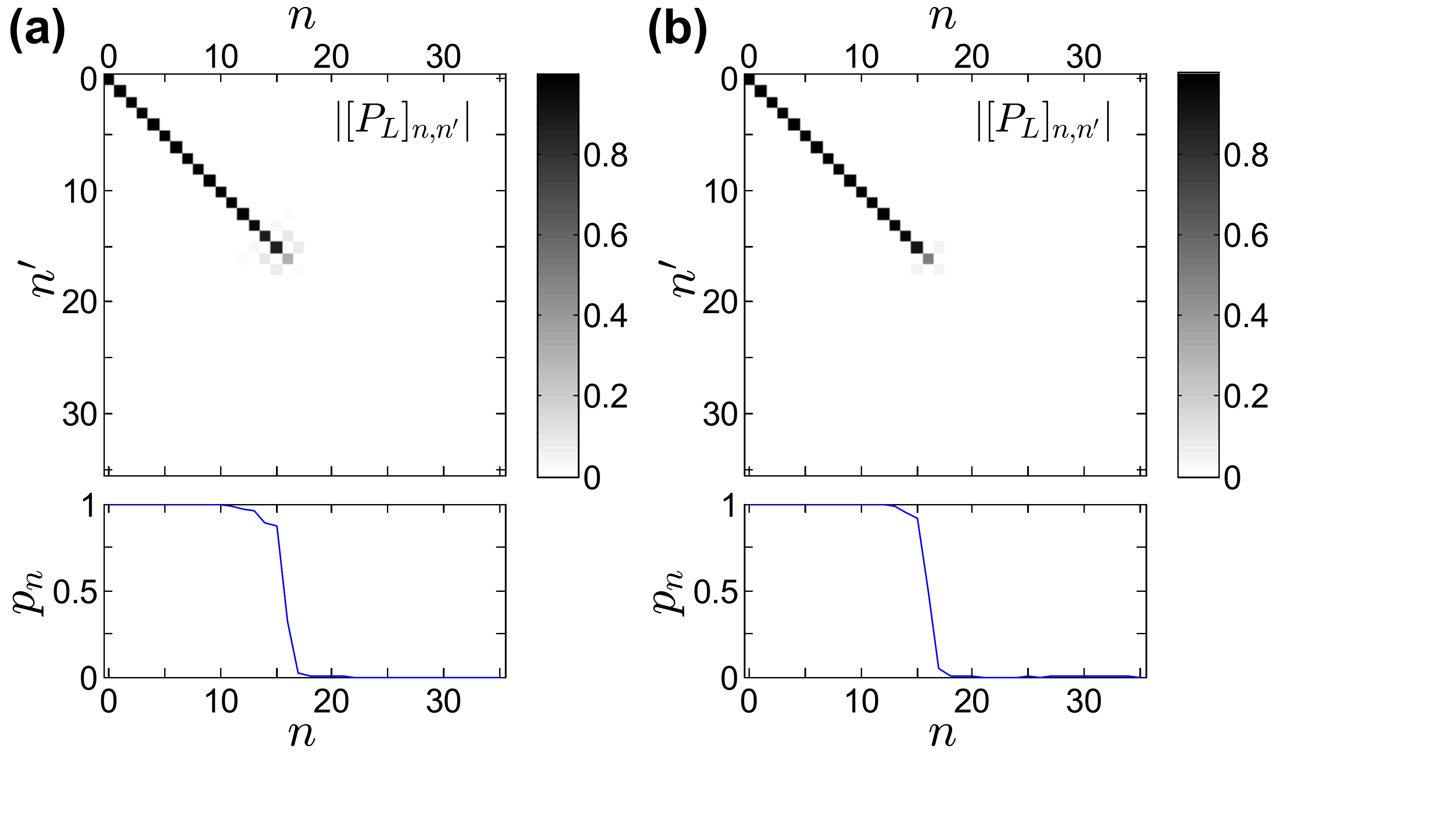}
  \caption{%Figure S1.
  $|[P_L]_{nn'}|$ (upper panels) and $p_n \equiv [P_L]_{nn}$ (lower panels) for (a) $\Lambda = 4$ and (b) $\Lambda = 8$. $L$ is chosen to be $\Lambda$ dependent as $L = k_\mr{F}^{-1} \Lambda^8$ for comparison; $n_L = 16$ for both of (a) and (b).
   Note that for $\Lambda = 4$, the largest off-diagonal element of $[P_L]_{nn'}$ is $\sim 0.07$ and the other off-diagonal ones are smaller by one or more orders. We choose $z = 0$.
  }
  \label{fig_prj}
  \end{center}
\end{figure}

\section{Partial trace over $x>L$} 
\label{TO}

%\section*{S3. Partial trace over $x>L$}
We develop a way of obtaining, from NRG state $\rho(T)$, the reduced density matrix $\rho(T,L)$ of the impurity and electrons in $x \leq L$, by tracing out states in $x > L$.  % with the aid of the correlation of energy and length scales of NRG states.
We use the projector to $x \leq L$,
\begin{equation}
  P_L = \frac{1}{a} \int_0^L \mr{d}x \prj{x}.
  \label{P_L}
\end{equation}
$\ket{x}$ is the state spatially localized at $x$ and $a$ is lattice constant. One has $\matel{\alpha n\sigma}{P_L}{\alpha' n'\sigma'} = \delta_{\alpha\alpha'}  \delta_{\sigma\sigma'} [P_L]_{nn'}$. The matrix $[P_L]_{nn'}$ is real symmetric and almost diagonal (see Fig.~\ref{fig_prj}). Its diagonal part $[P_L]_{nn}$ is finite for $n \lesssim n_L \equiv 2 \log_\Lambda k_\mr{F} L$, and vanishes for $n \gtrsim n_L$, reflecting the length scales of NRG sites. Off-diagonal parts $[P_L]_{nn'}$ are much smaller than diagonal ones, and decrease with increasing $\Lambda$, since spatial separation between $\ket{\alpha n \sigma}$'s increases. Based on this observation, we neglect the off-diagonal elements. The insensitivity of our computation of $\mc{E}_\mr{F}(\rho)$ to $\Lambda$ implies that this is a good approximation.

Neglecting the off-diagonal parts of $[P_L]_{nn'}$, we decompose $\ket{\alpha n\sigma}$ into the states of $x\leq L$ and $x>L$ as $f_{\alpha n\sigma}^\dagger  =  \sqrt{p_n} f_{\alpha n\sigma, \mr{in}}^\dagger + \sqrt{1-p_n} f_{\alpha n\sigma, \mr{out}}^\dagger$ where $p_n \equiv [P_L]_{nn}$.
%; the projected states created by $\{ f_{\alpha n\sigma, \mr{in}}^\dagger, f_{\alpha n\sigma, \mr{out}}^\dagger \}$ form an orthonormal basis set. 
Then the many-body occupation basis state $\ket{s_{\alpha n}}$ of site $n$ and channel $\alpha$ is expressed as
\begin{equation}
  \begin{aligned}
    \ket{0_{\alpha n}} &= \ket{0_{\alpha n}^\mr{out}} \ket{0_{\alpha n}^\mr{in}},  \\
    \ket{\ua_{\alpha n}} &= f_{{\alpha n}\ua}^\dagger \ket{0_{\alpha n}} \\
    &= ( \sqrt{p_n} f_{{\alpha n}\ua, \mr{in}}^\dagger + \sqrt{1-p_n} f_{{\alpha n}\ua, \mr{out}}^\dagger ) \ket{0_{\alpha n}^\mr{out}} \ket{0_{\alpha n}^\mr{in}}  \\
    &= \sqrt{p_n} \ket{0_{\alpha n}^\mr{out}} \ket{\ua_{\alpha n}^\mr{in}} + \sqrt{1-p_n} \ket{\ua_{\alpha n}^\mr{out}} \ket{0_{\alpha n}^\mr{in}},  \\
    \ket{\da_{\alpha n}} &= f_{{\alpha n}\da}^\dagger \ket{0_{\alpha n}} \\
    &= ( \sqrt{p_n} f_{{\alpha n}\da, \mr{in}}^\dagger + \sqrt{1-p_n} f_{{\alpha n}\da, \mr{out}}^\dagger ) \ket{0_{\alpha n}^\mr{out}} \ket{0_{\alpha n}^\mr{in}}  \\
    &= \sqrt{p_n} \ket{0_{\alpha n}^\mr{out}} \ket{\da_{\alpha n}^\mr{in}} + \sqrt{1-p_n} \ket{\da_{\alpha n}^\mr{out}} \ket{0_{\alpha n}^\mr{in}},  \\
    \ket{\ua\da_{\alpha n}} &= f_{{\alpha n}\ua}^\dagger f_{{\alpha n}\da}^\dagger \ket{0_{\alpha n}}  \\
    &= p_n \ket{0_{\alpha n}^\mr{out}} \ket{\ua\da_{\alpha n}^\mr{in}} + \sqrt{p_n (1-p_n)} \ket{\ua_{\alpha n}^\mr{out}} \ket{\da_{\alpha n}^\mr{in}}
    - \sqrt{p_n (1-p_n)} \ket{\da_{\alpha n}^\mr{out}} \ket{\ua_{\alpha n}^\mr{in}} + (1-p_n) \ket{\ua\da_{\alpha n}^\mr{out}} \ket{0_{\alpha n}^\mr{in}}.
  \end{aligned}
  \nonumber
\end{equation}
where $\ket{s_{\alpha n}^\mr{in (out)}}$ describes the occupation $s = 0, \ua, \da, \ua\da$ in the $x \leq L$ ($x>L$) part of site $n$ and channel $\alpha$. The Hilbert space of $x \leq L$ ($x>L$) is spanned by $\bigotimes_{\alpha;n} \{ \ket{s_\mr{\alpha n}^\mr{in (out)}} \}_s$. In this representation, tracing out $x > L$ is equivalent to partial trace over $\bigotimes_{\alpha;n} \{ \ket{s_\mr{\alpha n}^\mr{out}} \}_s$.

The partial trace over $\bigotimes_{\alpha;n} \{ \ket{s_\mr{\alpha n}^\mr{out}} \}_s$ can be efficiently done for $\rho(T)$ in Eq.~\eqref{rhoT} as $\rho (T,L) = \tr_0^\mr{out} \cdots \tr_N^\mr{out} \rho (T)$,
where $\tr_n^\mr{out} (\cdot ) \equiv \tr_{\alpha=1,n}^\mr{out} \cdots \tr_{\alpha=M,n}^\mr{out} (\cdot)$ is the partial trace applied to site $n$ and $\tr_{\alpha n}^\mr{out} (\cdot) \equiv \sum_s \epv{s_{\alpha n}^\mr{out}}{\cdot}$. %; here we utilize the fact that each $\rho_n$ involves the sites from 0 to $n$ and is decoupled from the sites $ > n$ as shown in Eq.~\eqref{rhoT}. 
By choosing appropriate basis states $\ket{{r}_{n i}} \in \mr{span} \bigotimes_{\alpha;n' \leq n} \{ \ket{s_{\alpha n'}^\mr{in}} \}_s$, we express $\rho (T,L)$, the result of the partial trace, as the block diagonal form of
\begin{equation}
%  \begin{gathered}
    \rho (T,L)  = \sum_n \rho_n^\mr{in} \otimes I_{> n}^\mr{in}, \quad
    \rho_n^\mr{in} = \sum_i {r}_{ni} \prj{{r}_{ni}}, \quad
    I_{>n}^\mr{in} = \bigotimes_{\alpha;n' > n} \sum_s \prj{s_{\alpha n'}^\mr{in}}.
%  \end{gathered}
  \nonumber 
\end{equation}
This form corresponds to Eq.~\eqref{rhoT} with $\rho_n \to \rho_n^\mr{in}$, $I_{> n} \to I_{> n}^\mr{in}$, and $\ket{E_{ni}^\mr{D}} \otimes \bigotimes_{\alpha; n' > n} \ket{s_{\alpha n'}} \to \ket{r_{ni}} \otimes \bigotimes_{\alpha; n' > n} \ket{s_{\alpha n'}^\mr{in}}$. $\textrm{Tr} ( \rho_n^\mr{in} \otimes I_{> n}^\mr{in} )$ is maximal near $n= \textrm{min} \{ n_T, n_L \}$.  As $\rho_n$'s are defined not spatially but energetically, $\rho_n^\mr{in}$ can be contributed from many $\rho_n$'s. 
In this form, we generalize the concept of the kept and discarded states used for $\rho(T)$ into $\rho(T,L)$. In each step of constructing $\rho_n^\mr{in}$, $\ket{{r}_{ni}}$'s are the ``discarded'' states with small $r_{ni}$, while there are the ``kept'' states %(with relatively low contribution in the construction) 
used for constructing $\rho_{n' > n}^\mr{in}$. 
  This form is useful for reducing the total number of basis states ($\ket{{r}_{ni}}$'s), hence, computation cost. It is similar to the truncation of density matrix renormalization group methods, and allows us to handle $\rho(T,L)$ with similar cost of computing $\rho (T)$.

%Note that the weight of each block $\tr ( \rho_n^0 \otimes I_{>n}^0 ) = \tr \rho_n^0 4^{M(N-n)}$ has a peak near $\min \{ n_T, n_L \}$; the inner bath sites $(\alpha, n' \gnsim n_T)$ are less occupied due to too small degeneracy factor and the sites $(\alpha, n' \gnsim n_L)$ tend to be traced out by the partial trace.

%Even in a two-qubit subspace, the witness for EoF is hard to construct directly since the witness does not have a simple form.
%\section{Derivation of Eq.~\eqref{X_i}}
\section{Derivation of two-qubit EW}
\label{XI}

We derive the optimal witness operator $X_\mr{2qb}$ for the EoF $\mc{E}_\mr{F} (\rho_\mr{2qb})$ of an arbitrary (unnormalized) two-qubit state $\rho_\mr{2qb}$ in the Hilbert space $\mc{H}_\mr{2qb} = \{ \ket{0}, \ket{1} \} \otimes \{ \ket{0}, \ket{1} \}$.
The derivation consists of two steps:
(i) Construct the EW $X^{\mc{C}}_\mr{2qb}$~\cite{Park10} for the concurrence~\cite{Wootters98} $\mc{C} (\rho_\mr{2qb})$, and
(ii) deduce $X_\mr{2qb}$ from $X^{\mc{C}}_\mr{2qb}$ by using the relation between the EoF and the concurrence;
for any normalized two-qubit state $\rho_\mr{2qb}/ \tr \rho_\mr{2qb}$, $\mc{E}_\mr{F}$ satisfies $\mc{E}_\mr{F} (\rho_\mr{2qb}/ \tr \rho_\mr{2qb}) = f (\mc{C} (\rho_\mr{2qb}/ \tr \rho_\mr{2qb}))$.
Here, $f(x) = h ( (1 + \sqrt{1 - x^2}) / 2 )$ and $h(x) = - x \log_2 x - (1-x) \log_2 (1-x)$.
Note that $\mc{E}_\mr{F}(\rho_\mr{2qb}/ \tr \rho_\mr{2qb}) = \mc{E}_\mr{F}(\rho_\mr{2qb}) / \tr \rho_\mr{2qb}$ and $\mc{C}(\rho_\mr{2qb}/ \tr \rho_\mr{2qb}) = \mc{C}(\rho_\mr{2qb}) / \tr \rho_\mr{2qb}$;
this type of relations holds for all convex-roof entanglement measures and beyond two qubits~\cite{Lee12}
and it is consistent with Eq.~(2). %Eq.~\eqref{OW}.
This is useful, as the ``two-qubit'' state $\rho_i = I_i \rho I_i$ in the main text is usually unnormalized. %Similarly, $\tr (X_{\rho_{2qb}} \rho_{2qb}/ \tr \rho_{2qb} ) = \mc{E}_\mr{F}(\rho_{2qb}/ \tr \rho_{2qb}) = \mc{E}_\mr{F}(\rho_{2qb}) / \tr \rho_{2qb} = \tr (X_{\rho_{2qb}} \rho_{2qb}) / \tr \rho_{2qb}$.

The derivation of $X_\mr{2qb}$ starts with the optimal witness operator $X^{\mc{C}}_\mr{2qb}$ for $\mc{C} (\rho_\mr{2qb})$. In the case of $\mc{C} (\rho_\mr{2qb}) \ne 0$, it is obtained~\cite{Park10,Lee12} as 
\begin{equation}
%  \begin{aligned}
    \mc{C} (\rho_\mr{2qb}) = \tr X^{\mc{C}}_\mr{2qb} \rho_\mr{2qb} 
    = \sup_{\mc{O}} \tr [\mc{O} ( 2 \prj{\Psi} - I_\mr{2qb} ) \mc{O}^\dagger \rho_\mr{2qb}], \quad
    \mc{O} = \mc{O}_1 \otimes \mc{O}_2, 
%  \end{aligned}
  \label{X^C}
\end{equation}
where $\mc{O}_i$ are local operators with determinant 1, each acting on $\{ \ket{0}, \ket{1} \}$.
Here $\ket{\Psi} = (\ket{0}\ket{0} + \ket{1}\ket{1}) / \sqrt{2}$ is maximally entangled ($\mc{E}_\mr{F} = 1$) Bell state.
In Eq.~\eqref{X^C}, $X_\mr{2qb}^{\mc{C}}$ is found by searching the optimal SLOCC (stochastic local quantum operations and classical communications in quantum information theory) operator $\mc{O}$ on $\mc{H}_\mr{2qb}$ that makes $\tr [\mc{O} ( 2 \prj{\Psi} - I_\mr{2qb} ) \mc{O}^\dagger \rho_\mr{2qb}]$ the largest.
The form of Eq.~\eqref{X^C} captures the invariance of $\mc{C}$ under SLOCC.
In the case of $\mc{C} (\rho_\mr{2qb}) = 0$, on the other hand, we choose $X^\mc{C}_\mr{2qb} = 0$ (the null operator).
$X^{\mc{C}}_\mr{2qb}$ provides $\mc{C} (\rho_\mr{2qb}) = \tr X_\mr{2qb}^\mc{C} \rho_\mr{2qb}$.
Searching the optimal operator $\mc{O}$ can be easily done by the singular value decomposition of the local operators as $\mc{O}_i = U_{1 i} F_{i} U_{2 i}$, where $U$'s are $2 \times 2$ local unitary operators and $F$ is a local filtering operator~\cite{Park10};
in the matrix representation, $F$ is written as $\left(\begin{smallmatrix} f & 0 \\ 0 & 1/f \end{smallmatrix} \right)$ with real $f$.
%Such local operator can be decomposed as $U_1 F U_2$ via singular value decomposition, where $U_1$ and $U_2$ are local unitary operators and $F$ is local filtering.
%For qubit, $F$ is simply written as $\left(\begin{smallmatrix} f & 0 \\ 0 & 1/f \end{smallmatrix} \right)$ for real $f$.
%When we apply SLOCC operators to quantum states, the normalization of them are not conserved generally, due to the existence of filtering operators.

$X_\mr{2qb}$ is obtained from $X^{\mc{C}}_\mr{2qb}$ and $\mc{E}_\mr{F} (\rho_\mr{2qb}/ \tr \rho_\mr{2qb}) = f (\mc{C} (\rho_\mr{2qb}/ \tr \rho_\mr{2qb}))$. As $f(x)$ is monotonically increasing and convex, one has $f (x) \geq f (x_\rho) + \left. \frac{\mr{d} f}{\mr{d} x} \right|_{x = x_\rho} ( x - x_\rho )$ at any $x_\rho$ and $x$. Substituting $x_\rho = \mc{C}(\rho_\mr{2qb} / \tr \rho_\mr{2qb})$, $f(x) \to X_\mr{2qb}$, $x \to X^{\mc{C}}_\mr{2qb}$, and $1 \to I_\mr{2qb}$, we choose
%\begin{equation}
%  f \left( \mc{C} \left( \frac{\rho'_i}{\tr \rho'_i} \right) \right) = \mc{E}_\mr{F} \left( \frac{\rho'_i}{\tr \rho'_i} \right) = \frac{\mc{E}_\mr{F} (\rho'_i)}{\tr \rho'_i} \geq f (x_i) + \left. \frac{\mr{d} f}{\mr{d} x} \right|_{x = x_i} \left( \mc{C} \left( \frac{\rho'_i}{\tr \rho'_i} \right) - x_i \right). \end{equation}
%Since $\tr \rho'_i > 0$, $\mr{d} f / \mr{d} x \geq 0$, and $\mc{C} (\rho'_i / \tr \rho'_i) = \mc{C} (\rho'_i) / \tr \rho'_i \geq \tr [ X_i^\mc{C} (\rho'_i / \tr \rho'_i)]$, we get
%\begin{equation}
%  \mc{E}_\mr{F} (\rho'_i) \geq f ( x_i ) \tr \rho'_i + \left. \frac{\mr{d} f}{\mr{d} x} \right|_{x = x_i} \left( \tr X_i^\mc{C} \rho'_i - x_i \tr \rho'_i \right), \end{equation}
%which implies that
\begin{equation}
%\begin{gathered}
 X_\mr{2qb} = f ( x_\rho ) I_\mr{2qb} + \left. \frac{\mr{d} f}{\mr{d} x} \right|_{x = x_\rho} \left( X^{\mc{C}}_\mr{2qb} - x_\rho I_\mr{2qb} \right), \quad
 x_\rho = \mc{C} ( {\rho_\mr{2qb}}/{ \tr \rho_\mr{2qb}} ) = {\mc{C}(\rho_\mr{2qb}) }/{ \tr \rho_\mr{2qb}}.
%\end{gathered}
 \label{X_2qb}
\end{equation}
This operator $X_\mr{2qb}$ is a witness operator for $\mc{E}_\mr{F}$, since for any state $\rho' \in \mc{H}_\mr{2qb}$, $\tr X_\mr{2qb} \rho' \le \mc{E}_\mr{F} (\rho')$;
one can check $\tr X_\mr{2qb} \rho' \le f(x_\rho) \tr \rho' + \frac{\mr{d}f}{\mr{d}x}|_{x=x_\rho} (\mc{C} (\rho') - x_\rho \tr \rho') \le f(\mc{C} (\rho' / \tr \rho')) \tr \rho' = \mc{E}_\mr{F} (\rho' / \tr \rho') \tr \rho' = \mc{E}_\mr{F} (\rho')$, using the convexity of $f(x)$ and Eq.~(2). %~\eqref{OW}.
Moreover, it is easy to show that $X_\mr{2qb}$ satisfies $\tr (X_\mr{2qb} \rho_\mr{2qb}) = \mc{E}_\mr{F} (\rho_\mr{2qb})$, using $\tr (X^\mc{C}_\mr{2qb} \rho_\mr{2qb}) = \mc{C} (\rho_\mr{2qb})$.
Therefore $X_\mr{2qb}$ is the optimal witness operator for $\mc{E}_\mr{F} (\rho_\mr{2qb})$.
%This proves Eq.~(5). %~\eqref{X_i}. 
Note that the elements of $\mbb{P}_{X_\mr{2qb}} = \{ \ket{\psi} \, | \, \mc{E}_\mr{F} ( \ket{\psi} ) = \epv{\psi}{X_\mr{2qb}} = \mc{E}_\mr{F} (\rho_\mr{2qb}) \}$ constitute the optimal pure-state decomposition for $\mc{E}_\mr{F} (\rho_\mr{2qb})$~\cite{Wootters98}.

\section{Witness operator $X = \sum_i X_i$}
\label{WO}

In the main text, we divide the whole Hilbert space $\mc{H}$ into ``two-qubit'' subspaces $\mc{H}_i \equiv \mr{span} \{ \ket{\eta} \otimes \ket{\phi_{i\eta'}} \}_{\eta \eta'}$ and obtain the optimal witness operator $X_i$ for $\mc{E}_\mr{F} (\rho_i)$, directly from $X_{i}$; $\rho_i = I_i \rho I_i$ is the projection of $\rho$ to $\mc{H}_i$ and $\tr X_i \rho_i = \mc{E}_\mr{F} (\rho_i)$. We here (i) prove that $X= \sum_i X_i$ is a witness operator for $\mc{E}_\mr{F} (\rho)$, namely that $\tr X \rho$ is a lower bound of $\mc{E}_\mr{F} (\rho)$, and also (ii) discuss a strategy how to optimize $X_i$'s.

The task (i) is equivalent, according to Eq.~(2), %~\eqref{OW}, 
to proving that $\tr (X \prj{\psi}) \le \mc{E}_\mr{F} (\ket{\psi})$ for any normalized pure state $\ket{\psi}$ in $\mc{H}$.
To prove it, we decompose $\ket{\psi} = \sum_i \ket{\psi_i}$, where $\ket{\psi_i}$ is the projection of $\ket{\psi}$ onto $\mc{H}_i$. 
Applying Schmidt decomposition to $\ket{\psi_i}$, $\ket{\psi_i} = c_{\Ua i} \ket{\Ua}\ket{\phi'_{i\Ua}} + c_{\Da i} \ket{\Da}\ket{\phi'_{i\Da}}$ and $\ovl{\phi'_{i \eta}}{\phi'_{i' \eta'}} = \delta_{ii'} \delta_{\eta \eta'}$, one has $\mc{E}_\mr{F} (\ket{\psi}) = h \left( \sum_i | c_{\Ua i} |^2 \right) = h \left( \sum_i | c_{\Da i} |^2 \right)$ where $h (x) = - x \log_2 x - (1-x) \log_2 (1-x)$. 
It satisfies $\mc{E}_\mr{F} (\ket{\psi}) \geq \sum_i \mc{E}_\mr{F} (\ket{\psi_i}) \geq \sum_i \epv{\psi_i}{X_i} = \tr (\sum_i X_i \prj{\psi})$; 
the first inequality is from the concavity of $h(x)$,
$h(|\sum c_{\Ua i} |^2) \geq$
$\sum_i (| c_{\Ua i} |^2 + | c_{\Da i} |^2) \, h ( \frac{| c_{\Ua i} |^2}{| c_{\Ua i} |^2 + | c_{\Da i} |^2} )$,
and also from $\mc{E}_\mr{F}(\ket{\psi_i}) = \langle \psi_i | \psi_i \rangle \mc{E}_\mr{F}(\ket{\psi_i}/\sqrt{\langle \psi_i | \psi_i \rangle})$, and the second from the fact that $X_i$ is a witness operator for the EoF of two-qubit states (which was proved above). 
This proves that $X$ is a witness operator for $\mc{E}_\mr{F} (\rho)$.

%As $h(x)$ is a concave function, we find $  \mc{E}_\mr{F} (\ket{\psi}) \geq \sum_i (| c_{\Ua i} |^2 + | c_{\Da i} |^2) \, h ( \frac{| c_{\Ua i} |^2}{| c_{\Ua i} |^2 + | c_{\Da i} |^2} )  = \sum_i (| c_{\Ua i} |^2 + | c_{\Da i} |^2) \, \mc{E}_\mr{F} ( \frac{\ket{\psi_i}}{\sqrt{\ovl{\psi_i}{\psi_i}}} )$.

Next, we discuss a strategy how to find $X = \sum_i X_i$ that provides a better lower bound of $\mc{E}_\mr{F} (\rho)$. One needs to first decompose $\mc{H}$ into $\mc{H}_i$'s. Among many possible ways for it, we choose a NRG-based way. In this way, we decompose $\rho$ into ``units'', and choose the basis state set $\mr{span} \{ \ket{\eta} \otimes \ket{\phi_{i\eta'}} \}_{i \eta \eta'}$ of each unit. Each unit has one or a few successive NRG diagonal blocks of $\rho$, and different units have no overlap; the number of blocks in a unit is chosen to have a better lower bound of $\mc{E}_\mr{F} (\rho)$. Then $\{ \ket{\Ua}, \ket{\Da} \} \otimes \{ \ket{\phi_{i\Ua}}, \ket{\phi_{i\Da}} \}$ constitutes $\mc{H}_i$. This way is naturally expected to lead to a good lower bound, as the NRG blocks capture the main physics.
After choosing $\mc{H}_i$'s, we find the optimal witness $X_i$ (equivalently $X^{\mc{C}}_{i}$) for $\mc{E}_\mr{F} (\rho_i) = \tr X_i \rho_i$, following Eqs.~\eqref{X^C} and \eqref{X_2qb}.
%\begin{equation}
%  \begin{gathered}
%    X_i =  f ( x_0 ) I_i + \left. \frac{\mr{d} f}{\mr{d} x} \right|_{x = x_0} \left( X^{\mc{C}}_{\rho_i} - x_0 I_i \right), \quad x_0 = \mc{C}\left(\frac{\rho_i}{ \tr \rho_i}\right) = \frac{\mc{C}(\rho_i) }{ \tr \rho_i},  \\
%  \mc{C} (\rho_i) = \tr X^{\mc{C}}_{\rho_i} \rho_i = \sup_{\mc{O}_i} \tr [\mc{O}_i ( 2 \prj{\psi^+_i} - I_i ) \mc{O}^\dagger_i \rho_i],  \quad \mc{O}_i = \mc{O}_{i, \eta} \otimes \mc{O}_{i,\phi}, 
%  \end{gathered}
%  \label{X_i2}
%\end{equation}
%where $\rho_i = I_i \rho I_i$ and $\ket{\psi^+_i} = (\ket{\Ua} \ket{\phi_{i\Ua}} + \ket{\Da} \ket{\phi_{i\Da}})/\sqrt{2}$. To find it, one searches the optimal operator $\mc{O}_i$ that leads to $\sup_{\mc{O}_i} \cdots$ in Eq.~\ref{X_i2}.
We skip other technical details of finding $X$, such as how to choose the basis states $ \{ \ket{\phi_{i\eta}} \}_{i \eta}$.

\section{Upper bound of $\mc{E}_\mr{F} (\rho)$}
\label{UB}

In the above, we find $X = \sum_i X_i$ that provides the best lower bound of $\mc{E}_\mr{F} (\rho)$ within the form utilizing Eq.~\eqref{X_2qb}. We call this operator as $X_\rho^\mr{opt}$. A good upper bound is also obtained from $X_\rho^\mr{opt}$, by finding a set of pure states $\mbb{P}_{X_\rho^\mr{opt}} = \{ \ket{\psi} \, | \, \epv{\psi}{X_\rho^\mr{opt}} = \mc{E}_\mr{F} ( \ket{\psi} ) \}$ and a decomposition $\rho = \sum_l p_l' \prj{\psi_l'}$ where each $\ket{\psi_l'}$ is sufficiently similar to an element of $\mbb{P}_{X_\rho^\mr{opt}}$. We suggest below a systematic way of finding the decomposition $\rho = \sum_l p_l' \prj{\psi_l'}$.

To find the decomposition, we diagonalize $\rho = \sum_d \bar{p}_d \prj{\bar{\psi}_d}$, where $\ovl{\bar{\psi}_{d}}{\bar{\psi}_{d'}} = \delta_{dd'}$.
Any pure-state decomposition $\rho = \sum_l q_l \prj{\varphi_l}$ is generated by a left-unitary matrix $U$ as $\sqrt{q_l} \ket{\varphi_l} = \sum_d U_{l d} \sqrt{\bar{p}_d} \ket{\bar{\psi}_d}$ and $U^\dagger U = I$.
To generate $\{ \ket{\varphi_l} \}$ close to $\mbb{P}_{X_\rho^\mr{opt}} = \{ \ket{\psi_{l}} \}$, 
we introduce a matrix $W$, $[W]_{ld} = \ovl{\bar{\psi}_d}{\psi_l} p_l^{y_1} \bar{p}_d^{y_2}$, and obtain its singular value decomposition of $W = V_\mr{L} \Sigma V_\mr{R}^\dagger$, where $y_1$ and $y_2$ are the variables to be optimized. Here, $p_l$'s are chosen to satisfy $\rho \simeq \sum_l p_l \prj{\psi_l}$. Then, we choose $U$ as $U = V_\mr{L} V_\mr{R}^\dagger$, and use it to obtain $\{ \ket{\psi'_l} \}$ via $\sqrt{p'_l} \ket{\psi'_l} = \sum_d U_{l d} \sqrt{\bar{p}_d} \ket{\bar{\psi}_d}$. Finally, we optimize $y_1$ and $y_2$ to minimize $\sum_l p'_l \mc{E}_\mr{F} (\ket{\psi'_l})$. The minimum value of $\sum_l p'_l \mc{E}_\mr{F} (\ket{\psi'_l})$ is a good upper bound of $\mc{E}_\mr{F} (\rho)$.

%note that $W$ is left-unitary when $\{ \ket{\psi_l'} \} \subset \mbb{P}_{X_\rho^{opt}}$ and $y_1 = -y_2 = 1/2$. 

%Fortunately, in virtue of the construction $X = \sum_i X_i$, the majority of $\mbb{P}_X$ can be identified as $\{ \ket{\psi_{ij}} \}_{ij} = \bigcup_i \{ \ket{\psi_{ij}} \}_{j}$ where $\{ \ket{\psi_{ij}} \}_{j}$ are the pure states in the optimal pure-state decomposition of $\rho_i$ for EoF, i.e., $\rho_i = \sum_j p_{ij} \prj{\psi_{ij}}$ and $\mc{E}_\mr{F} (\rho_i) = \sum_j p_{ij} \mc{E}_\mr{F} (\ket{\psi_{ij}})$ \cite{Wootters98}. Note that $\ket{\psi_{ij}} \in \mbb{P}_{X_i}$. Hereafter, for simplicity, we combine the indices $l \equiv (i, j)$. If $\rho = \sum_l p_l \prj{\psi_{l}}$ holds, we assure the global optimality of $X$. It does not hold, however, for general situations.

In the above way of finding a upper bound, it takes heavy numerical cost to handle $\rho$ as a whole, since $\rho$ has a large size. To avoid the heavy cost, we decompose $\rho = \sum_n \rho_n$ into the NRG blocks $\rho_n$'s (or the units of a few successive blocks), construct a witness operator $X_n$ for $\mc{E}_\mr{F} (\rho_n)$, and find a good upper bound $\mc{E}_n$ of $\mc{E}_\mr{F} (\rho_n)$, using $\mbb{P}_{X_n}$ as mentioned above.
The sum $\sum_n \mc{E}_n$ of the upper bound of $\mc{E}_\mr{F} (\rho_n)$ over $n$'s provides a good upper bound of $\mc{E}_\mr{F} (\rho)$. Note that 
%the decomposition $\rho = \sum_n \rho_n$ is different from the decomposition $\mc{H} = \bigoplus \mc{H}_i$, and that 
$\sum_n X_n$ is not necessarily a witness operator of $\rho$; it is because $X_n$ is not necessarily constructed by the bath states orthogonal between different NRG blocks (or units), contrary to $X = \sum_i X_i$.% in Eq.~\ref{X_i2}.

\section{$T$ dependence of $\mc{E}_\mr{F}$ from bosonization}
\label{TB}

We here confirm the universal power-law thermal decay of $\mc{E}_\mr{F}$,
%our numerical result of the thermal power-law decay of $\mc{E}_\mr{F}$ at $L \to \infty$ and $T \ll T_{1CK,2CK}$ in Eq.~\eqref{K_T}, 
using finite-size bosonization and refermionization methods~\cite{Zarand00}, and attribute the power-law exponents different between 1CK and 2CK to Majorana fermions emerging in 2CK.

For 1CK and 2CK, the thermal state has the form of $\rho = \sum_i w_i \prj{E_i}$;
$w_i$ is Boltzmann weight. $\ket{E_i}  = b_{i \Ua} \ket{\Ua} \ket{e_{i\Ua}} + b_{i \Da} \ket{\Da} \ket{e_{i\Da}}$ is an energy eigenstate with energy $E_i$ and an eigenstate of the total (impurity and bath) spin-$z$ operator simultaneously. Bath states $\ket{e_{i \eta}}$ satisfy $\langle e_{i \Ua} | e_{i \Da} \rangle = 0$ because $\ket{e_{i \Ua}}$ and $\ket{e_{i \Da}}$ have different spin-$z$ quantum numbers, while $\langle e_{i \eta} | e_{i' \ne i \eta'} \rangle \ne 0$ in general.
We focus on $\ket{E_i}$'s with $E_i \sim k_\mr{B} T$, as they govern the properties of $\rho$; this is due to the competition between degeneracy and Boltzmann weight.
Using the bosonization, we will later show that for $E_i, E_{i'} \sim k_\mr{B} T \ll k_\mr{B} T_\mr{1CK,2CK}$,  $S_{z, i i'} \equiv \matel{E_i}{S_z}{E_{i'}}$ and $S_{-, i i'} \equiv \matel{E_i}{S_-}{E_{i'}}$ satisfy
\begin{equation}
  S_{z, i i'}, \,\, S_{-, i i'} \,\, \propto
  \begin{cases}
    T / T_\mr{1CK}, & \text{ for 1CK,} \\
    \sqrt{T / T_\mr{2CK}}, & \text{ for 2CK.} \\
  \end{cases}
  \label{Delta_Sz_S-}
\end{equation}
The $S_z$ and $S_-$ impurity spin operator, $S_{z} \equiv (\prj{\Ua} - \prj{\Da}) / 2$ and $S_{-} \equiv | \Da \rangle \langle \Ua |$, have entanglement information. $S_{z, i i}$ connects with $\mc{E}_\mr{F} (\ket{E_i})$. $\ket{E_i}$ is maximally entangled when $S_{z, i i} = (|b_{i \Ua}|^2 - |b_{i \Da}|^2)/2 = 0$, while it is separable when $S_{z, i i} = \pm 1/2$.
We find that for $E_i /k_\mr{B} \sim T \ll T_\mr{1CK,2CK}$, $|S_{z, ii}| \ll 1/2$ and $\mc{E}_\mr{F} (\ket{E_i}) = h \left( \frac{1}{2} + S_{z, ii} \right) \simeq 1 - 2 |S_{z, ii}|^2 /\log 2$, where $h(x) = -x \log_2 x - (1-x) \log_2 (1-x)$. On the other hand, $S_{z, i i' \ne i}$ and $S_{-, i i' \ne i}$ have the information of state overlap $\langle e_{i\eta}| e_{i'\eta'} \rangle$.  From Eq.~\eqref{Delta_Sz_S-} and $\langle E_i | E_{i'} \rangle= \delta_{i i'}$, we find 
\begin{equation}
  \ovl{e_{i\eta}}{e_{i'\eta'}} - \delta_{ii'} \delta_{\eta \eta'} \propto
  \begin{cases}
    T / T_\mr{1CK}, & \text{ for 1CK,} \\
    \sqrt{T / T_\mr{2CK}}, & \text{ for 2CK.} \\
  \end{cases}
  \label{e_overlap}
\end{equation}
The overlap results in entanglement reduction in a pure-state mixture, $\mc{E}_\mr{F} ( \sum_i w_i \prj{E_i} ) \leq \sum_i w_i \mc{E}_\mr{F} (\ket{E_i})$. From Eq.~(1), %~\eqref{ConvRoof},  
$\mc{E}_\mr{F,0} \equiv \sum_i w_i \mc{E}_\mr{F} (\ket{E_i}) \simeq \sum_i w_i (1 - 2 |S_{z, ii}|^2 / \log 2)$ is a upper bound of $\mc{E}_\mr{F} (\rho)$. The upper bound $\mc{E}_\mr{F,0}$ and Eq.~\eqref{Delta_Sz_S-} agree with the power law in Eq.~(4). %~\eqref{K_T}.  

We also confirm Eq.~(4) %~\eqref{K_T} 
using a lower bound of $\mc{E}_\mr{F} (\rho)$. We consider a witness operator $X'$,
\begin{equation}
  \begin{gathered}
    X' = \sum_{i} \left[ \frac{2}{\log 2} \prj{\Psi_{i}} - \left( \frac{2}{\log 2} - 1 \right) I_{i} \right], \\
    \ket{\Psi_{i}} = \frac{1}{\sqrt{2}} (\ket{\Ua}\ket{\phi_{i \Ua}} + \ket{\Da}\ket{\phi_{i \Da}}), \quad
    I_{i} = \sum_{\eta =\Ua, \Da; \eta' =\Ua, \Da} \prj{\eta} \otimes \prj{\phi_{i \eta'}}.
  \end{gathered} \label{Xprime}
\end{equation}
This has the similar form to Eq.~(3). %~\eqref{X1}.  
Here, $\ket{\phi_{i\eta}}$'s are the orthonormal states obtained by applying the Gram-Schmidt orthonormalization process to the states $\{ \ket{e_{i\eta}} \}$. Because $\langle e_{i\eta}| e_{i'\eta'} \rangle - \delta_{i i'} \delta_{\eta \eta'}$ is very small at $T \ll T_\mr{1CK,2CK}$ as in Eq.~\eqref{e_overlap}, $\ket{\phi_{i\eta}}$ little deviates from $\ket{e_{i\eta}}$ as $b_{i\eta} \ket{e_{i\eta}} = (\ket{\phi_{i\eta}} + \ket{\delta_{i\eta}})/\sqrt{2}$. The expectation value $\tr X' \rho$ is a lower bound of $\mc{E}_\mr{F} (\rho)$. After some computation, we find 
\begin{equation}
%\begin{aligned}
  \tr X' \rho = 1 - \frac{1}{2 \log 2} \sum_{i i'} w_i \large[ | \ovl{\phi_{i'\Ua}}{\delta_{i\Ua}} - \ovl{\phi_{i'\Da}}{\delta_{i\Da}} |^2 
  + 2 ( | \ovl{\phi_{i'\Da}}{\delta_{i\Ua}} |^2 + | \ovl{\phi_{i'\Ua}}{\delta_{i\Da}} |^2 ) \large].
%\end{aligned}
     \nonumber
\end{equation}
Applying $\ket{\delta_{i \eta}} \propto T/T_\mr{1CK}$ for 1CK and $\ket{\delta_{i \eta}} \propto \sqrt{T/T_\mr{2CK}}$ for 2CK in Eq.~\eqref{e_overlap}, we find that $\tr X' \rho$ satisfies Eq.~(4). %~\eqref{K_T}.
This analytic derivation of the same universal power-law behavior of the upper and lower bounds $\mc{E}_{\mr{F},0}$ and $\tr X' \rho$ strongly supports our numerical result of Eq.~(4). %~\eqref{K_T}.
% Note that the numerically results in Fig.~\ref{fig_res} are better than $\mc{E}_\mr{F,0}$ and $\tr X' \rho$. 

For a complete proof, we now derive Eq.~\eqref{Delta_Sz_S-} and discuss the difference between 1CK and 2CK. 
We first consider 2CK. 
In 2CK, the electron bath has the four degrees of freedom, total charge, total spin, charge difference between the channels, and spin difference between the channels. 
According to the bosonization and refermionization along Emery-Kivelson line~\cite{Zarand00}, 
the degree of freedom from the spin difference (decoupled from the others) is described by the resonant-level model, $H^\mr{2CK}_x = \epsilon_\mr{d} {\,:\,}c_\mr{d}^\dagger c_\mr{d}{\,:\,} + \sum_k \epsilon_k {\,:\,}c_k^\dagger c_k{\,:} + \sqrt{\Delta \Gamma} \, \sum_k (c_k^\dagger + c_k) (c_\mr{d} - c_\mr{d}^\dagger)$, 
where $c_\mr{d}^\dagger$ creates a pseudofermion in the resonant level coupled to a reservoir of pseudofermions (with momentum $k$ and energy $\epsilon_k$) created by $c_k^\dagger$, $\Delta$ is the level spacing of the reservoir, and $: \, :$ means normal ordering.
$\Gamma$ is the broadening of the resonance and plays the role of Kondo temperature, $\Gamma = k_\mr{B} T_\mr{2CK}$. We choose $\Delta$ as $\Delta = k_\mr{B} T$ to focus on energy scale $\sim k_\mr{B} T$.
$c_\mr{d}^\dagger$ ($c_\mr{d}$) corresponds to impurity spin raising operator $S_+$ (lowering $S_-$), while $c^\dagger_\mr{d} c_\mr{d} = S_z + 1/2$.
In our case of no external magnetic field, $\epsilon_\mr{d} =0$, and Majorana fermion $\gamma_\mr{d+} = (c_\mr{d} + c_\mr{d}^\dagger)/\sqrt{2}$ decouples from $H_x^\mr{2CK}$ (while the other Majorana $\gamma_\mr{d-} = i(c_\mr{d}^\dagger - c_\mr{d})/\sqrt{2}$ participates in $H_x^\mr{2CK}$).
Namely,  a  half of the impurity decouples from bath electrons, making 2CK a non-Fermi liquid.
$H^\mr{2CK}_x$ is diagonalized as $H^\mr{2CK}_x = \sum_{\epsilon \geq 0} \epsilon\, c_{2\epsilon}^\dagger c_{2\epsilon} + \sum_{k > k_\mr{F}} \epsilon_k  d_k^\dagger d_k + (\text{const.})$.
Meanwhile, each of other three degrees of freedom is bosonic and diagonalized as $\sum_q q b_{q y}^\dagger b_{q y}$, where $b_{q y}$ is a bosonic operator for the degree of freedom $y$ with momentum $q = n_q \Delta / \hbar v_\mr{F}$ and $n_q$ is a positive integer.
The eigenstates $\ket{\tilde{E}^\mr{2CK}_i}$ of the refermionized Hamiltonian are the direct products of the eigenstates of $H_x^\mr{2CK}$ and the eigenstates of the three bosonic degrees of freedom.
%and its eigenstates are denoted as ; $\ket{\tilde{E}^\mr{2CK}_i}$ also has the other quantum numbers decoupled from $H_x^\mr{2CK}$ such as total charge/spin and the charge difference. 

We compute $S_{z, i i'} =  \matel{E^\mr{2CK}_i}{S_z}{E^\mr{2CK}_{i'}}$. The eigenstates $\ket{E^\mr{2CK}_{i}}$ of 2CK connects with the eigenstates $ \ket{\tilde{E}^\mr{2CK}_i} = U_\mr{EK} \ket{E^\mr{2CK}_{i}} $  via Emery-Kivelson transformation $U_\mr{EK}$~\cite{Zarand00}. 
Since $U_\mr{EK} S_z U_\mr{EK}^\dagger = S_z = c_\mr{d}^\dagger c_\mr{d} - 1/2 = i \gamma_{\mr{d}+} \gamma_{\mr{d} -}$, $S_{z, i i'}$ is written as $S_{z, i i'} = i \matel{\tilde{E}^\mr{2CK}_i}{\gamma_{\mr{d}+} \gamma_{\mr{d} -}}{\tilde{E}^\mr{2CK}_{i'}} $. After some calculations, we find 
\begin{equation}
  \begin{aligned}
    S_{z, i i'} &= i \matel{\tilde{E}^\mr{2CK}_i}{\gamma_{\mr{d}+} \gamma_{\mr{d} -}}{\tilde{E}^\mr{2CK}_{i'}} \\
    &= \frac{1}{2} \sum_{\epsilon, \epsilon' \geq 0} B_{\epsilon \mr{d} +} B_{\epsilon' \mr{d} -}  \matel{\tilde{E}^\mr{2CK}_i}{(c_{2\epsilon} + c_{2 \epsilon}^\dagger) (c_{2\epsilon'}^\dagger - c_{2\epsilon'})}{\tilde{E}^\mr{2CK}_{i'}} \\
    &= \frac{1}{2} \sum_{\epsilon' \geq 0} B_{\epsilon' \mr{d} -}  \matel{\tilde{E}^\mr{2CK}_i}{(c_{2 \epsilon=0}^\dagger + c_{2 \epsilon=0})  (c_{2\epsilon'}^\dagger - c_{2\epsilon'})}{\tilde{E}^\mr{2CK}_{i'}},
  \end{aligned} \nonumber
\end{equation}
where coefficient $B_{\epsilon \mr{d} \pm}$ connects $\gamma_{\mr{d} \pm}$ and the excitation of $(c_{2\epsilon -}^\dagger \pm c_{2\epsilon -})/\sqrt{2}$ and $|\matel{\tilde{E}^\mr{2CK}_i}{(c_{2 \epsilon=0}^\dagger + c_{2 \epsilon=0})  (c_{2\epsilon'}^\dagger - c_{2\epsilon'})}{\tilde{E}^\mr{2CK}_{i'}}| = 1$ or 0; for the detail of $B_{\epsilon \mr{d} \pm}$, see Ref.~\cite{Zarand00}.  
In the last equality,  we used $B_{\epsilon \mr{d} +} = \delta_{\epsilon 0}$, coming from the decoupling of Majorana fermion $\gamma_{\mr{d}+}$ from the bath. Since $B_{\epsilon \mr{d} -} \propto \sqrt{T/T_\mr{2CK}}$ at $T \ll T_\mr{2CK}$, $S_{z, i i'} \propto \sqrt{T/T_\mr{2CK}}$ in agreement with Eq.~\eqref{Delta_Sz_S-}. 

We also compute $S_{-, i i'} =  \matel{E^\mr{2CK}_i}{S_-}{E^\mr{2CK}_{i'}}$. Using $U_\mr{EK}$ and $c_\mr{d} = F^\dagger_s S_-$, where $F^\dagger_s$ is a Klein factor, we have $S_{-, i i'} = \matel{\tilde{E}^\mr{2CK}_i}{e^{-i \varphi_s (0)} F_s c_\mr{d}}{\tilde{E}^\mr{2CK}_{i'}}$, where the boson field $\varphi_s (0)$ results from the commutation between $S_-$ and $U_\mr{EK}$; see Ref.~\cite{Zarand00}.
$\varphi_s$ and $F_s$ correspond to total charge degree of freedom.
Here, $F_s$ gives  1 or 0, hence, not related with $T/T_\mr{2CK}$.
And, $c_\mr{d} = (c_{2 \epsilon = 0}^\dagger + c_{2 \epsilon = 0})/2 + O (\sqrt{T/ T_\mr{2CK}})$ does not provide $\sqrt{T/ T_\mr{2CK}}$ in the leading order term.
In contrast, $e^{-i \varphi_s (0)}$ interestingly provides $e^{-i \varphi_s (0)} \propto \sqrt{T/ T_\mr{2CK}}$, since the bosonic reservoir, included in the resonant-level model as being decoupled from $H_x^\mr{2CK}$, also has the finite length of $\sim h v_\mr{F} / \Delta \sim h v_\mr{F} / k_\mr{B} T$.
We show this, expanding $\varphi_s (0)$ in terms of boson operators $b_{q s}$, $\varphi_s (0) = \sum_{q > 0} \frac{-1}{\sqrt{n_q}} (b_{q s}^\dagger + b_{q s}) e^{-aq/2}$, where $a \propto \Gamma^{-1}$ is the cutoff. 
Some calculations lead to
\begin{equation}
%\begin{aligned}
e^{-i \varphi_s (0)} = \prod_{q > 0} \left[ \exp \left( i \frac{1}{\sqrt{n_q}} b_{q s}^\dagger e^{-aq/2} \right) 
     \exp \left( i\frac{1}{\sqrt{n_q}} b_{q s} e^{-aq/2} \right) \,\, \exp \left( - \frac{1}{2n_q} e^{-aq} \right) \right] . 
%\end{aligned}
    \nonumber
\end{equation}
The first and second terms in the squared bracket are $O(1)$ since $\ket{\tilde{E}_i^\mr{2CK}}$ are eigenstates of $b_{q s}^\dagger b_{q s}$ with eigenvalues 0 or 1.
Meanwhile,
$\prod_{q>0} \exp \left( - \frac{1}{2n_q} e^{-aq} \right) = \sqrt{1 - e^{- a \Delta / \hbar v_\mr{F}}} \propto \sqrt{a \Delta }= \sqrt{T / T_\mr{2CK}}$ at $T \ll T_\mr{2CK}$. Hence, $S_{-, i i'} \propto \sqrt{T / T_\mr{2CK}}$ is proved. 
%in agreement with Eq.~\eqref{Delta_Sz_S-}.

Next, we derive Eq.~\eqref{Delta_Sz_S-} for 1CK.
According to the bosonization and refermionization at Toulouse point~\cite{Zarand00},
the spin degree of freedom of 1CK is also described by a similar resonant-level model, $H^\mr{1CK}_s = \epsilon_\mr{d} {\,:\,}c_\mr{d}^\dagger c_\mr{d}{\,:\,} + \sum_k \epsilon_k {\,:\,}c_k^\dagger c_k{\,:} + \sqrt{\Delta \Gamma} \, \sum_k (c_k^\dagger c_\mr{d} + c_\mr{d}^\dagger c_k) = \sum_{\epsilon} \epsilon {\,:\,}c_{1\epsilon}^\dagger c_{1\epsilon}{\,:} + (\text{const.})$, but with $\Gamma = k_\mr{B} T_\mr{1CK}$.
Contrary to 2CK, it shows a Fermi liquid, and no Majorana fermion of the impurity decouples from the bath.
We compute $S_{z, i i'} =  \matel{E^\mr{1CK}_i}{S_z}{E^\mr{1CK}_{i'}}$, where $\ket{\tilde{E}^\mr{1CK}_i}$'s denote the eigenstates of $H^\mr{1CK}_s$.
Using another Emery-Kivelson transformation $U_\mr{EK}^\mr{1CK}$, $ \ket{\tilde{E}^\mr{1CK}_i} = U_\mr{EK}^\mr{1CK} \ket{E^\mr{1CK}_{i}} $,
we find $S_{z, ii'} = \matel{E^\mr{1CK}_i}{S_z}{E^\mr{1CK}_{i'}} = \matel{\tilde{E}^\mr{1CK}_i}{c_\mr{d}^\dagger c_\mr{d}}{\tilde{E}^\mr{1CK}_{i'}} - \delta_{i i'}/2$,
since $U_\mr{EK}^\mr{1CK} S_z (U_\mr{EK}^\mr{1CK})^\dagger = S_z = c_\mr{d}^\dagger c_\mr{d} - 1/2$.
It is written as $ S_{z, ii'} = \sum_{\epsilon \epsilon'} B_{\epsilon \mr{d}} B_{\epsilon' \mr{d}} \matel{\tilde{E}^\mr{1CK}_i}{c_{1\epsilon}^\dagger c_{1\epsilon'}}{\tilde{E}^\mr{1CK}_{i'}} - \delta_{ii'} / 2$ in terms of the coefficients $B_{\epsilon \mr{d}}$ connecting $c_\mr{d}$ and $c_{1 \epsilon}$.
Since $B_{\epsilon \mr{d}} \propto \sqrt{T/T_\mr{1CK}}$ at $T \ll T_\mr{1CK}$~\cite{Zarand00} and $\matel{\tilde{E}^\mr{1CK}_i}{c_{1\epsilon}^\dagger c_{1\epsilon'}}{\tilde{E}^\mr{1CK}_{i'}} =1$ or 0, we find $S_{z, i i' \ne i} \propto T / T_\mr{1CK}$, in agreement with  Eq.~\eqref{Delta_Sz_S-}.
Similarly, it is straightforward to show $S_{z, i i} \propto T / T_\mr{1CK}$.

We also compute $S_{-,i i'} = \matel{E_i^\mr{1CK}}{S_-}{E_{i'}^\mr{1CK}}$. The bosonization results in an expression similar to the 2CK, $S_{-,i i'} = \matel{\tilde{E}^\mr{1CK}_i}{e^{-i (\sqrt{2}-1) \varphi_s (0)} e^{-i\pi S_z} c_\mr{d}}{\tilde{E}^\mr{1CK}_{i'}}$. It is however hard to handle $e^{-i (\sqrt{2}-1) \varphi_s (0)}$ with the irrational number $\sqrt{2}-1$ of Toulouse point. Instead, we study $S_{-,i i'}$ using an effective theory near the strong-coupling fixed point~\cite{Hewson93}.
At the fixed point, the Kondo singlet state decouples from Fermi-liquid excitations.
Near the fixed point at $T \ll T_\mr{1CK}$, the singlet and the excitations are coupled, with coupling energy $\sim D \Lambda^{-(N-1)/4} \sim \sqrt{T}$. This modifies $\ket{E_i^\mr{1CK}}$ from $\ket{E_i^\mr{1CK, 0}}$ as $\ket{E_i^\mr{1CK}} = \ket{E_i^\mr{1CK, 0}} + \ket{\delta_i}$, where $\ket{E_i^\mr{1CK, 0}}$'s are the states at the fixed point. The coupling energy leads to $\langle \delta_i | \delta_i \rangle \propto T$, resulting in $S_{-,i i'} \propto T$, in agreement with Eq.~\eqref{Delta_Sz_S-}. The same argument reproduces $S_{z, i i'} =  \matel{E^\mr{1CK}_i}{S_z}{E^\mr{1CK}_{i'}} \propto T / T_\mr{1CK}$, which was obtained using the bosonization in the above.

%$|G_\mr{1CK} \rangle = 2^{-1/2} (\ket{\Ua} f_{0 \downarrow}^\dagger - \ket{\Da} f_{0 \uparrow}) |\mr{vacuum} \rangle$ decouples with Fermi-liquid excitations $\tilde{\epsilon}_k$ ($< T \ll T_{1CK}$) created by $f_{n > 0 \sigma}^\dagger$, where $f_{n \sigma}^\dagger$ creates an electron in the $n$-th NRG chain. The singlet state has the energy $\propto T_{1CK}$. Near the fixed point at low temperature, the excited states are modified by the coupling between the singlet state and the states by $f_{n > 0 \sigma}^\dagger$, and the coupling strength $\propto \sqrt{T}$. The modification is computed by the first-order perturbation on the bare energies $T_{1CK}$ and $\tilde{\epsilon}_k$ by the coupling strength $T$. This results in $\matel{E^\mr{1CK}_i}{S_-}{E^\mr{1CK}_{i'}} \propto T / (T_{1CK} - \tilde{\epsilon}_k) \simeq T/T_{1CK}$, which agrees with Eq.~\eqref{Delta_Sz_S-}.

\section{$L$ dependence of $\mc{E}_\mr{F}$ for 1CK}
\label{L1CK}

Our numerical result of the $L$ dependence of $\mc{E}_\mr{F}$ at $T=0$ and $L \gg \xi_\mr{1CK}$ in Eq.~(5) %~\eqref{1CK_L}
is reproduced with the variational 1CK ground state by Yosida~\cite{Yosida66},
\begin{equation}
%\begin{gathered}
\ket{\psi_\mr{Y}} = \frac{\ket{\Ua} \ket{\phi_{\mr{Y} \da}} - \ket{\Da} \ket{\phi_{\mr{Y} \ua}}}{\sqrt{2}}, \quad
  \ket{\phi_{\mr{Y} \sigma}} = \phi_{\mr{Y} \sigma}^\dagger \ket{\mr{FS}}, \quad
  \phi_{\mr{Y} \sigma}^\dagger = \frac{1}{\sqrt{\mc{N}}} \sum_{k > k_\mr{F}} \frac{c_{k\sigma}^\dagger}{\epsilon_k + E_\mr{Y}},
%\end{gathered}
  \nonumber % \label{psiY}
\end{equation}
where $\ket{\mr{FS}} = \prod_{k \leq k_\mr{F}; \sigma = \ua,\da} c_{k \sigma}^\dagger \ket{0}$ is the Fermi sea of the bath, $\ket{0}$ is the vacuum state, $\mc{N}$ is the normalization factor ensuring $|\langle \phi_{\mr{Y} \sigma}| \phi_{\mr{Y} \sigma} \rangle |^2 = 1$, and $E_\mr{Y} = D e^{-4/3J\nu_\mr{F}}$ corresponds to $k_\mr{B} T_\mr{1CK}$; we here use $c_{k \sigma}^\dagger \equiv c_{\alpha =1, k \sigma}^\dagger$.
This illustrates the Kondo singlet of the impurity spin and the electron spin created by $\phi_{\mr{Y}\sigma}^\dagger$.
The spatial dependence of $\phi_{\mr{Y}\sigma}^\dagger$ is
$\phi_\mr{Y} (x) = \frac{1}{\sqrt{\mc{N}}} \sum_{k > k_\mr{F}} \sqrt{\frac{2}{l}} \frac{\sin k x}{\epsilon_k + E_\mr{Y}}$, where $l \to \infty$ is the total length of the one-dimensional bath and $E_\mr{Y} \equiv \hbar v_\mr{F} k_\mr{Y} \simeq \hbar v_\mr{F} / \xi_\mr{1CK}$.

%\begin{equation}
%  \phi_\mr{Y} (x) = \frac{1}{\sqrt{\mc{N}}} \sum_{k > k_\mr{F}} \sqrt{\frac{2}{l}} \frac{\sin k x}{\epsilon_k + E_\mr{Y}} \simeq \sqrt{\frac{2}{\pi k_\mr{Y}}} \frac{\cos k_\mr{F} x}{x} \,\,\,\,\, \text{ for } x \gg k_\mr{F}^{-1}, \,\, k_\mr{Y}^{-1} \nonumber
%\end{equation}

To study the $L$ dependence of $\mc{E}_\mr{F}$, we compute $\tr_{x>L} \prj{\psi_\mr{Y}}$, by tracing out the states outside $L$. For this purpose, we decompose each single-electron operator,
\begin{equation}
%\begin{gathered}
c_{k\sigma}^\dagger = \sqrt{\frac{L}{l}} \, c_{k \sigma, \mr{in}}^\dagger + \sqrt{1 - \frac{L}{l}} \, c_{k \sigma, \mr{out}}^\dagger, \quad
  \phi_{\mr{Y}\sigma}^\dagger = \sqrt{1-p} \, \phi_{\mr{Y} \sigma, \mr{in}}^\dagger + \sqrt{p} \, \phi_{\mr{Y} \sigma, \mr{out}}^\dagger,
%\end{gathered}
 \nonumber % \label{cksigma_phiysigma}
\end{equation}
where $c^\dagger_{k\sigma,\mr{in (out)}} \sim \int_{x \leq L (x>L)} \mr{d}x \, c^\dagger_{x \sigma}  \sin kx$ creates an electron inside (outside) $L$ and  $\phi^\dagger_{\mr{Y}\sigma, \mr{in (out)}} \sim \int_{x \leq L (x>L)} \mr{d}x \, c^\dagger_{x \sigma} \phi_{\mr{Y} \sigma}(x)$ ($c_{x\sigma}^\dagger$ creates an spin-$\sigma$ electron at $x$).
$p = \int_L^\infty \mr{d}x \, | \phi_\mr{Y} (x) |^2 \simeq 1 / \pi k_\mr{Y} L \simeq \xi_\mr{1CK} / \pi L$ is the probability of finding the electron of $\phi_\mr{Y}$ outside $L$. Accordingly, the Fermi sea is written as $\ket{\mr{FS}} = \prod_{\substack{k\leq k_\mr{F} \\ \sigma = \ua,\da}} \left( \sqrt{\frac{L}{l}} \, c_{k \sigma, \mr{in}}^\dagger + \sqrt{1-\frac{L}{l}} \, c_{k \sigma, \mr{out}}^\dagger \right) \ket{0}_\mr{in} \ket{0}_\mr{out} \simeq_{ \lim_{l \rightarrow \infty}} \ket{0}_\mr{in} \ket{\mr{FS}}_\mr{out}$,
%\begin{equation}
%  \ket{\mr{FS}} = % \prod_{\substack{k\leq k_\mr{F} \\ \sigma = \ua,\da}} c_{k\sigma}^\dagger \ket{0}
%  \prod_{\substack{k\leq k_\mr{F} \\ \sigma = \ua,\da}} \left( \sqrt{\frac{L}{l}} \, c_{k \sigma, \mr{in}}^\dagger + \sqrt{\frac{l - L}{l}} \, c_{k \sigma, \mr{out}}^\dagger \right) \ket{0}_\mr{in} \ket{0}_\mr{out} 
% \simeq_{ \lim_{l \rightarrow \infty}} \ket{0}_\mr{in} \ket{\mr{FS}}_\mr{out}, \nonumber \end{equation}
where $\ket{0}_\mr{in (out)}$ denotes the vacuum state of $x \leq L$ ($x > L$) and $\ket{\mr{FS}}_\mr{out} \equiv \prod_{\substack{k\leq k_\mr{F} \\ \sigma = \ua,\da}} c_{k \sigma, \mr{out}}^\dagger \ket{0}_\mr{out}$ is the Fermi sea outside $L$. Here, we used $l \gg L$, where the portion of plane waves inside $L$ can be ignored and $\ket{\mr{FS}}_\mr{out}$ is well defined. Using the decomposition, we find
\begin{equation}
%\begin{aligned}
\ket{\psi_\mr{Y}} \simeq \frac{1}{\sqrt{2}} \left[ \ket{\Ua} (\sqrt{1-p} \, \phi_{\mr{Y} \da, \mr{in}}^\dagger + \sqrt{p} \, \phi_{\mr{Y} \da, \mr{out}}^\dagger) 
- \ket{\Da} (\sqrt{1-p} \, \phi_{\mr{Y} \ua, \mr{in}}^\dagger + \sqrt{p} \, \phi_{\mr{Y} \ua, \mr{out}}^\dagger) \right] \ket{0}_\mr{in} \ket{\mr{FS}}_\mr{out}.
%\end{aligned}
   \nonumber
\end{equation}
Then, we compute
$\tr_{x > L} \prj{\psi_\mr{Y}} = \sum_i \ovl{\psi_{i, \mr{out}}}{\psi_\mr{Y}} \ovl{\psi_\mr{Y}}{\psi_{i, \mr{out}}}$,
where $\ket{\psi_{i, \mr{out}}}$'s are relevant states outside $L$, $\ket{\psi_{i, \mr{out}}} \in \{ \ket{\mr{FS}}_\mr{out}, \phi_{\mr{Y} \ua, \mr{out}}^\dagger \ket{\mr{FS}}_\mr{out}, \phi_{\mr{Y} \da, \mr{out}}^\dagger \ket{\mr{FS}}_\mr{out} \}$. The result is
%So the resulting mixed state becomes rank-3,
\begin{equation}
%\begin{aligned}
  \tr_{x>L} \prj{\psi_\mr{Y}} \simeq (1 - p) \prj{\psi_{1,\mr{in}}}
  + \frac{p}{2} \prj{\psi_{2,\mr{in}}} + \frac{p}{2} \prj{\psi_{3,\mr{in}}},
%\end{aligned}
  \label{TrPsiY} 
\end{equation}
where $\ket{\psi_{1,\mr{in}}} \equiv \frac{1}{\sqrt{2}} ( \ket{\Ua} \phi_{\mr{Y} \da, \mr{in}}^\dagger - \ket{\Da} \phi_{\mr{Y} \ua, \mr{in}}^\dagger ) \ket{0}_\mr{in}$, $\ket{\psi_{2,\mr{in}}} \equiv \ket{\Da} \ket{0}_\mr{in}$, and $ \ket{\psi_{3,\mr{in}}} \equiv \ket{\Ua} \ket{0}_\mr{in}$. 

We calculate $\mc{E}_\mr{F}(\tr_{x>L} \prj{\psi_\mr{Y}})$, using a witness operator similar to Eq.~(3), %~\eqref{X1}, 
\begin{equation}
  X_\mr{Y} = \frac{2}{\log 2} \prj{\psi_{1,\mr{in}}} - \left(\frac{2}{\log 2} - 1\right) I_{1,\mr{in}},
  \label{X_Y}
\end{equation}
where $I_{1,\mr{in}} = \sum_{\eta = \Ua, \Da; \sigma = \uparrow, \downarrow} \prj{\eta} \otimes \prj{\phi_{\mr{Y}\sigma, \mr{in}}}$ and $\ket{\phi_{\mr{Y} \sigma, \mr{in}}} = \phi_{\mr{Y} \sigma, \mr{in}}^\dagger \ket{0}_\mr{in}$. 
This operator is the optimal witness operator for $\mc{E}_\mr{F}(\ket{{\psi_{i,\mr{in}}}})$ with $i=1,2,3$, namely, it provides the exact value of $\mc{E}_\mr{F} (\ket{{\psi_{i=1,2,3,\mr{in}}}})$; 
one checks $\mc{E}_\mr{F} (\ket{{\psi_{1,\mr{in}}}}) = \epv{{\psi_{1,\mr{in}}}}{X_\mr{Y}} = 1$, $\mc{E}_\mr{F} (\ket{{\psi_{2,\mr{in}}}}) = \mc{E}_\mr{F} (\ket{{\psi_{3,\mr{in}}}}) = 0$. 
According to the duality~\cite{Lee12,Ryu12} between Eqs.~(1) and~(2). %~\eqref{ConvRoof} and~\eqref{OW},
the expectation value of $X_\mr{Y}$ equals the exact value of $\mc{E}_\mr{F}$ for any mixture of $\ket{{\psi_{i=1,2,3,\mr{in}}}}$ including $\mc{E}_\mr{F}(\tr_{x>L} \prj{\psi_\mr{Y}})$. 
We obtain $\mc{E}_\mr{F}(\tr_{x>L} \prj{\psi_\mr{Y}}) = \tr [ X_\mr{Y} ( \tr_{x>L} \prj{\psi_\mr{Y}} ) ] \simeq 1 - p$, namely, $1 - \mc{E}_\mr{F}(\tr_{x>L} \prj{\psi_\mr{Y}}) \simeq p \propto \xi_\mr{1CK} / L$. 
This confirms the universal power law in Eq.~(5), %\eqref{1CK_L},
which we numerically find in the main text. This computation based on $X_\mr{Y}$ indicates the usefulness of witness operators for analytically studying macroscopic entanglement EoF in many-body mixed-states.

%\section*{Supplementary Note}

\section{Kondo cloud at finite temperature}
\label{TL}

\begin{figure}[bt]
\begin{center}
  \includegraphics[width=0.7\textwidth]{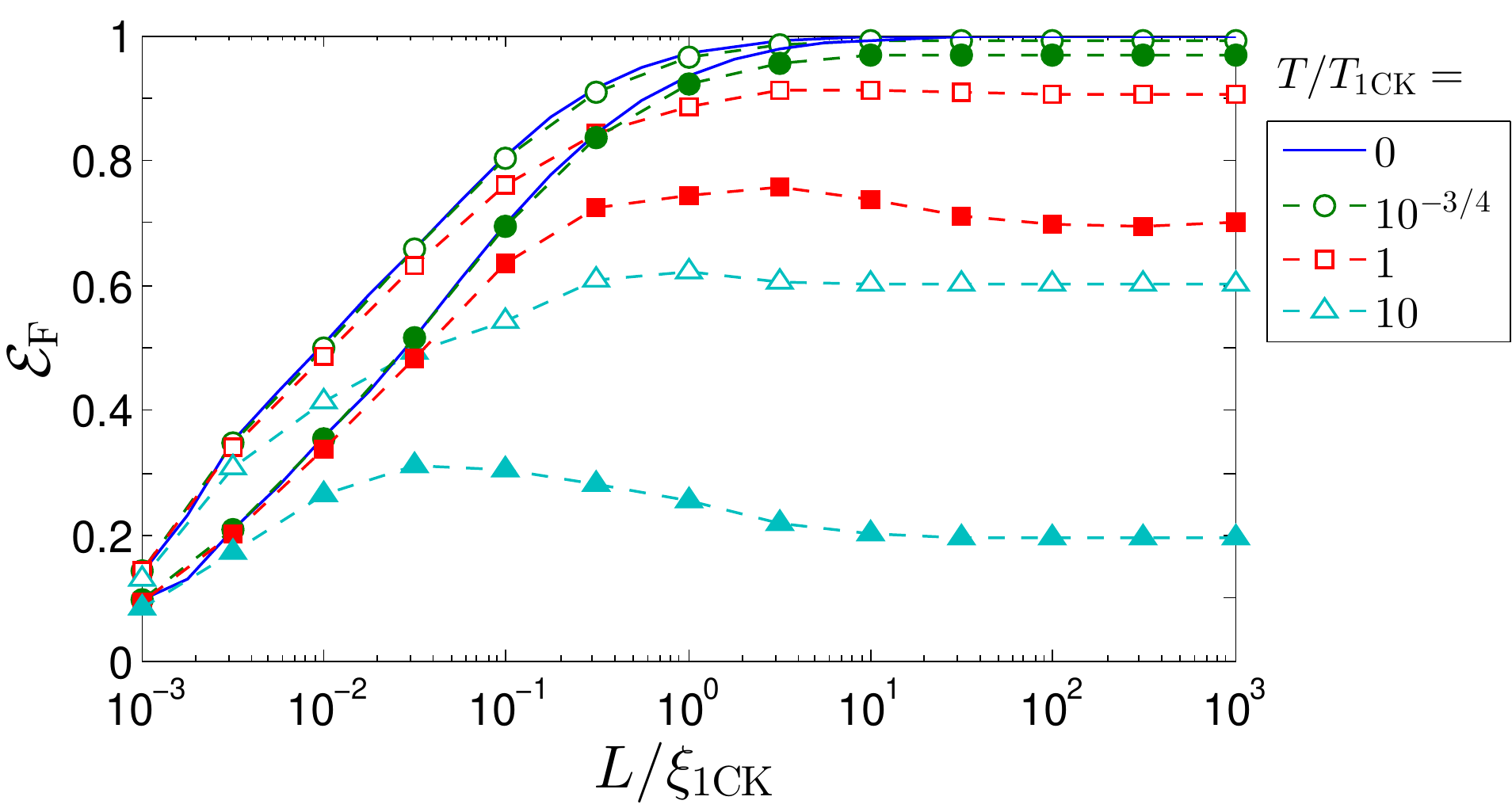}
  \caption{%Figure S2.
  Kondo cloud at finite temperature. Dependence of EoF $\mc{E}_\mr{F}$ on $L$ at different $T$'s, $T/T_\mr{1CK} = 0, 10^{-3/4} \, (\simeq 0.18), 1, 10$; the results of $T/T_\mr{1CK} = 0$ and $10^{-3/4}$ are almost overlapped. This shows that the cloud size is about $\xi_\mr{1CK}$ at $T \lesssim T_\mr{1CK}$, while it decreases at $T \gtrsim T_\mr{1CK}$ as $T$ increases. %roughly as $\hbar v_F / k_\mr{B} T$. 
  Empty (filled) symbols represent a upper (lower) bound of $\mc{E}_\mr{F}$.
  }
  \label{fig_cloud}
  \end{center}
\end{figure}

\begin{figure}[bt]
\begin{center}
  \includegraphics[width=0.6\textwidth]{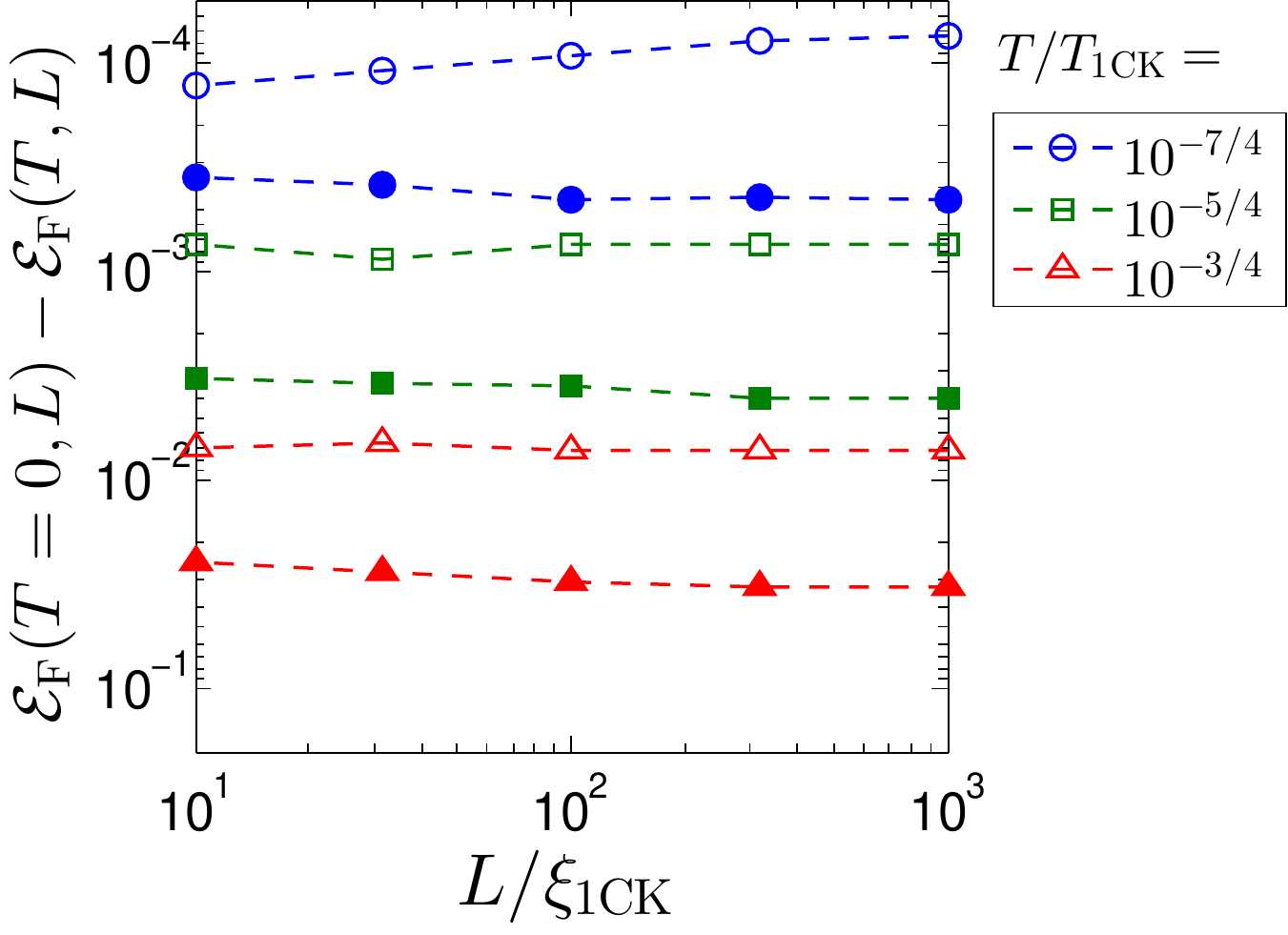}
  \caption{%Figure S3. 
  Dependence of $\mc{E}_\mr{F} (T = 0) - \mc{E}_\mr{F} (T)$ on $L$ at different $T$'s, $T / T_\mr{1CK} = 10^{-3/4} \, (\simeq 0.18), 10^{-5/4} \, (\simeq 0.056) , 10^{-7/4} \, (\simeq 0.018)$. This shows that $\mc{E}_\mr{F} (T = 0) - \mc{E}_\mr{F} (T)$ is almost independent of $L$ at $T \ll T_\mr{1CK}$ and $L \gg \xi_\mr{1CK}$. Empty (filled) symbols represent a upper (lower) bound of $\mc{E}_\mr{F}$.
  }
  \label{fig_cloud2}
  \end{center}
\end{figure}

In Fig.~\ref{fig_cloud}, we present our numerical result of the dependence of $\mc{E}_\mr{F}(\rho)$ on $L$ at finite $T$. Figure~\ref{fig_cloud} shows that as $L$ decreases, $\mc{E}_\mr{F}$ starts to decrease near $L \simeq \xi_\mr{1CK}$ at $T \lesssim T_\mr{1CK}$, while roughly near thermal length $L \simeq L_T \equiv \hbar v_\mr{F} / k_\mr{B} T$ at $T \gtrsim T_\mr{1CK}$. 
This means that the size of Kondo cloud is $\xi_\mr{1CK}$ and robust against thermal effects at $T \lesssim T_\mr{1CK}$, while it is roughly $L_T$, decreasing with increasing $T$, at $T \gtrsim T_\mr{1CK}$.
Moreover, Fig.~\ref{fig_cloud2} suggests that the two 1CK power-law decays in Eqs.~(4) and~(5) %~\eqref{K_T} and~\eqref{1CK_L}
are additive at $T \ll T_\mr{1CK}$ and $L \gg \xi_\mr{1CK}$,
\begin{equation}
\mc{E}_\mr{F} \simeq 1 - a_1 \left(\frac{T}{T_\mr{1CK}}\right)^2 - b_1 \left( \frac{\xi_\mr{1CK}}{L} \right). \label{K_TL}
\end{equation}
Together with the fact that the two 1CK power laws are not connected by the usual replacement of $k_\mr{B} T \leftrightarrow \hbar v_\mr{F} / L$  by the uncertainty relation (as their power-law exponents are different), these unusual findings indicate that the mechanism of entanglement suppression by thermal effects differs from that by the partial trace over $x>L$. Note that we are unable to definitely conclude whether the cloud size is $L_T$ at $T \gtrsim T_\mr{1CK}$, because the numerical results of the upper and lower bounds of $\mc{E}_\mr{F}$ are not close enough to each other; the witness operator $X$ is devised from the entanglement feature of the ground and low-energy eigenstates, hence, less efficient at $T \gtrsim T_\mr{1CK}$ or $L \lesssim \xi_\mr{1CK}$.

All these findings can be understood by the following argument. At finite $T$, $\mc{E}_\mr{F} (\rho)$ is mainly contributed by the excited states $\ket{E_i}  = b_{i \Ua} \ket{\Ua} \ket{e_{i\Ua}} + b_{i \Da} \ket{\Da} \ket{e_{i\Da}}$ of $E_i \sim k_\mr{B} T$. They have $\mc{E}_\mr{F} (\ket{E_i}) \simeq 1 - 2 |S_{z, ii}|^2 /\log 2 = 1 - (|b_{i \Ua}|^2 - |b_{i \Da}|^2)^2/ (2 \log 2)$. 
For larger $E_i$, $|S_{z, ii}|^2 \propto (|b_{i \Ua}|^2 - |b_{i \Da}|^2)^2$ increases, as $\ket{E_i}$ more deviates from the exact Bell state. % $\ket{G}$.
Our numerical results imply that the dependence of $\mc{E}_\mr{F}$ on $T$ reflects this behavior, hence, the entanglement of excited states $\ket{E_i}$.
On the other hand, the $L$ dependence of $\mc{E}_\mr{F}$ is related to the loss of the wave functions of $\ket{e_{i\Ua}}$ and $\ket{e_{i\Da}}$ by the partial trace over $x>L$.
At $T \ll T_\mr{1CK}$ and $L \gg \xi_\mr{1CK}$, the two mechanisms ($|S_{z, ii}|^2$ and the partial wave-function loss) seem to work independently,
resulting in the additive scaling law in Eq.~\eqref{K_TL} as $\mc{E}_\mr{F} \simeq (1 - a_1 (T/T_\mr{1CK})^2)(1 - b_1 \xi_\mr{1CK}/L) \simeq 1 - a_1 (T/T_\mr{1CK})^2 - b_1 \xi_\mr{1CK}/L$.
The size of Kondo cloud, measured by $\mc{E}_\mr{F}$, may directly reflect the spatial extension of the wave functions $\langle x | e_{i \Ua} \rangle$ and $\langle x | e_{i \Da} \rangle$ participating in excited-state entanglement. 
%This agrees with the NRG result that $\langle x | e_{i \Ua} \rangle$ and $\langle x | e_{i \Da} \rangle$ extend mainly within $\xi_{1CK}$ at $T \lesssim T_{1CK}$. % and over $L_T$ at $T \gtrsim T_{1CK}$.

\end{widetext}

\end{document}